\documentclass[acmsmall,screen]{acmart}

\AtBeginDocument{%
  \providecommand\BibTeX{{%
    \normalfont B\kern-0.5em{\scshape i\kern-0.25em b}\kern-0.8em\TeX}}}

\setcopyright{acmlicensed}
\copyrightyear{2025}
\acmYear{2025}
\acmDOI{XXXXXXX.XXXXXXX}

\acmConference[ISSTA '25]{Proceedings of the 2025 ACM SIGSOFT International Symposium on Software Testing and Analysis}{June 25--28, 2025}{Trondheim, Norway}

\acmISBN{978-1-4503-XXXX-X/18/06}

\makeatletter
\let\@authorsaddresses\@empty
\makeatother

\usepackage{amsmath,amsfonts}
\usepackage{algorithm}
\usepackage{algorithmic}
\usepackage{booktabs}
\usepackage{color}
\usepackage{cleveref}
\usepackage[commandnameprefix=ifneeded,final]{changes}
\usepackage{enumitem}
\usepackage{framed}
\usepackage{graphicx}
\usepackage{hyperref}
\usepackage{listings}
\usepackage{mdframed}
\usepackage[many]{tcolorbox}
\usepackage{xcolor}
\usepackage{xspace}

\newcommand{\toolname}{\textsc{Doctor}\xspace}

\DeclareRobustCommand{\mybox}[2][gray!20]{%
\begin{tcolorbox}[
        breakable,
        left=0pt,
        right=0pt,
        top=0pt,
        bottom=0pt,
        colback=#1,
        colframe=#1,
        width=\linewidth, 
        enlarge left by=0mm,
        boxsep=5pt,
        arc=0pt,outer arc=0pt,
        ]
        #2
\end{tcolorbox}
}

\usepackage[font=small,skip=2pt]{caption}
\newcommand{\distance}{2pt}
\begin{document}

\title{\toolname: Optimizing Container Rebuild Efficiency by Instruction Re-Orchestration}

\author{Zhiling Zhu}
\affiliation{%
  \institution{Zhejiang University of Technology}
  \city{Hangzhou}
  \country{China}
}
\orcid{0009-0004-0859-0129}
\email{zhilingzhu@zjut.edu.cn}

\author{Tieming Chen}
\authornote{Tieming Chen and Chengwei Liu are the corresponding authors.}
\affiliation{%
  \institution{Zhejiang University of Technology}
  \city{Hangzhou}
  \country{China}
}
\orcid{xxxx}
\email{tmchen@zjut.edu.cn}

\author{Chengwei Liu}
\authornotemark[1]
\affiliation{%
  \institution{Nanyang Technological University}
  \city{Singapore}
  \country{Singapore}
}
\orcid{xxxx}
\email{chengwei.liu@ntu.edu.sg}

\author{Han Liu}
\affiliation{%
  \institution{The Hong Kong University of Science and Technology}
  \city{Hong Kong}
  \country{China}
}
\orcid{xxx}
\email{liuhan@ust.hk}

\author{Qijie Song}
\affiliation{%
  \institution{Zhejiang University of Technology}
  \city{Hangzhou}
  \country{China}
}
\orcid{xxxx}
\email{songqijie@zjut.edu.cn}

\author{Zhengzi Xu}
\affiliation{%
  \institution{Nanyang Technological University}
  \city{Singapore}
  \country{Singapore}
}
\orcid{0000-0002-8390-7518}
\email{zhengzi.xu@ntu.edu.sg}

\author{Yang Liu}
\affiliation{%
  \institution{Nanyang Technological University}
  \city{Singapore}
  \country{Singapore}
}
\orcid{xxxx}
\email{yangliu@ntu.edu.sg}

\renewcommand{\shortauthors}{Zhu, et al.}

\begin{abstract}

Containerization has revolutionized software deployment, with Docker leading the way due to its ease of use and consistent runtime environment. As Docker usage grows, optimizing Dockerfile performance, particularly by reducing rebuild time, has become essential for maintaining efficient CI/CD pipelines. However, existing optimization approaches primarily address single builds without considering the recurring rebuild costs associated with modifications and evolution, limiting long-term efficiency gains. To bridge this gap, we present \toolname, a method for improving Dockerfile build efficiency through instruction re-ordering that addresses key challenges: identifying instruction dependencies, predicting future modifications, ensuring behavioral equivalence, and managing the optimization's computational complexity. We developed a comprehensive dependency taxonomy based on Dockerfile syntax and a historical modification analysis to prioritize frequently modified instructions. \added{Using a weighted topological sorting algorithm, \toolname optimizes instruction order to minimize future rebuild time while maintaining functionality. Experiments on 2,000 GitHub repositories show that \toolname improves 92.75\% of Dockerfiles, reducing rebuild time by an average of 26.5\%, with 12.82\% of files achieving over a 50\% reduction. Notably, 86.2\% of cases preserve functional similarity.} These findings highlight best practices for Dockerfile management, enabling developers to enhance Docker efficiency through informed optimization strategies.

\end{abstract}




\maketitle

\section{Introduction}

The rapid adoption of containerization has revolutionized software deployment, with Docker becoming a key technology due to its ease of use and consistent runtime environment. Recent reports estimate the container market will reach USD 15.06 billion by 2028~\cite{container-market}. Containerization allows applications to bundle dependencies and configurations within isolated environments that share host resources, providing a more lightweight and efficient alternative to traditional virtualization~\cite{watada2019emerging, bentaleb2022containerization, rad2017introduction}.

However, as development teams increasingly rely on Docker, optimizing Dockerfile performance, particularly reducing rebuild time, has become a pressing concern. Dockerfiles, which define how containers are built, often contain inefficiencies that can lead to prolonged rebuild times, significantly impacting the development lifecycle. Excessive rebuild times not only slow down development but also increase resource consumption and hinder continuous integration and delivery (CI/CD) pipelines, where swift rebuilds are critical for rapid feedback. Common inefficiencies in Dockerfiles, such as inappropriate layer ordering~\cite{dockerfile-order}, redundant commands~\cite{dockerfile-redundant-commands}, and missed caching opportunities~\cite{dockerfile-better-use-cache}, contribute to these delays. For instance, placing frequently updated instructions at the top of the Dockerfile can prevent the Docker cache from being effectively utilized, leading to repeated rebuilds of downstream layers. 
However, considering the continuous maintenance and evolution activities, deficiency, such as excessive storage demands (i.e., large image layers) ~\cite{zhao2019large, lin2020large}, Dockerfile code smells ~\cite{wu2020characterizing, rosa2024fixing, durieux2024empirical}, and outdated dependencies ~\cite{zerouali2021multi, zerouali2019relation}, is unavoidable.

To improve build efficiency, research has been conducted to investigate solutions for the optimization of docker build performance.
Huang et al. ~\cite{huang2019fastbuild} proposed FastBuild, a caching method that intercepts and caches remote file requests, achieving a 10x speed increase and substantial data reduction. Zhao et al. ~\cite{zhao2020large} and Durieux ~\cite{durieux2024empirical} have also conducted studies to investigate and find solutions to enhance storage efficiency and reduce Docker smells. However, these methods primarily target single builds, and none of the existing work has taken the recurring rebuild costs associated with ongoing modifications and project evolution into consideration, limiting their overall efficiency improvements. To this end, we aim to find out possible optimizations for Dockerfile that consider not only the single build efficiency but also the overall efficiency of docker rebuild in future maintenance activities.

To bridge this gap, we still face the following challenges: 
\textbf{C1: Inner Dependencies in Dockerfile.} Considering that Dockerfiles are piled up by a series of Dockerfile Instructions~\cite{shabanimirzaei2024dockerfile}, there could be explicit or implicit dependencies that follow-up instructions would rely on components, variables, or settings executed by previous instructions. However, there are no existing taxonomies or tools for the identification of these dependencies. Moreover, Docker has defined its own Domain Specific Language (DSL) for the parsing and processing of Dockerfile, and there is no official DSL schema released for public usage, which makes it more challenging to precisely identify the dependencies. 
\textbf{C2: Future Modification Prediction.} Since we aim to optimize Dockerfiles towards their efficiency in future rebuilds, which instructions in the Dockerfiles could be more frequently modified should be considered in the optimization target function, while no existing solutions have been given.
\textbf{C3: Behavior Equivalence of Optimization.} It is also crucial to ensure the equivalence of Dockerfile behavior during the Docker build to ensure its compatibility with its original user requirements.
\textbf{C4: State Space Explosion in Optimization Algorithm.} The number of possible instruction permutations grows exponentially with Dockerfile length, posing a significant challenge in efficiently identifying optimal solutions that honor dependency constraints, it is also non-trivial to ensure the computation of optimization feasible for large and complex Dockerfiles.

To address these challenges, we propose \toolname, a comprehensive approach to enhancing Dockerfile build efficiency through instruction re-ordering. Specifically, 
For \textbf{C1}, we first revisited the official documentation of Dockerfile, and proposed an Extended Backus-Naur Form (EBNF) presentation for Dockerfile grammar, based on which, we implemented a robust parser for instruction interpretation and identified the key elements that may introduce dependency relations in dockerfile. Based on that, we also propose a comprehensive classification of Dockerfile inner dependency for dependency constraint identification.
For \textbf{C2}, we referred to the historical modification of Dockerfiles to investigate their possibility of future modifications. Specifically, considering that instructions could be altered for different purposes, it is difficult to distinguish modification and removal, we adopted similarity analyses with different measurement strategies to identify possible modifications, and aggregate the historical records to approach the future modification possibility. 
For \textbf{C3}, we strictly followed the identified dependency relations among instructions to retain the dependencies of groups of instructions, and optimized instruction orders with these partial relationships.
For \textbf{C4}, we adopted a weighted topological sorting algorithm to optimize the instruction orders by minimizing its total build time with their future modification possibility considered. To avoid state space explosion, we greedily prioritized the instructions that have the largest occupation of building time cost in future rebuilds during the topological sorting algorithm. 

\added{Our experiments demonstrate the effectiveness and efficiency of \toolname on 2,000 randomly selected popular GitHub repositories. \toolname improves 92.75\% of Dockerfiles in the dataset, reducing future rebuild time by an average of 26.5\%, with 12.82\% of Dockerfiles achieving a reduction of over 50\%. On average, \toolname takes only 77.55 seconds to optimize each Dockerfile.} Moreover, \toolname also demonstrated an excellent performance on the preservation of original Dockerfile functionalities during the optimization. Our experiments also showed that 86.2\% Dockerfiles still produced images with the same directory structures (including file-system, environment variables, package manager installations, and WORKDIR), and most of the rest still retained functional similarity. All unit test cases from 23 filtered repositories remained passed after optimization. Only 0.21\% of Dockerfiles exhibited semantic differences after manual inspection. 
Based on these results, we also concluded the patterns that contributed the most to the optimization of Dockerfiles, which developers could further practice to guide their Dockerfile management. 

In summary, the main contributions of this paper are as follows:

\begin{itemize}[leftmargin=*]
    \item We propose \toolname, a novel and comprehensive approach for optimizing Dockerfile rebuild efficiency. To the best of our knowledge, \toolname is the first tool to extract dependencies between Dockerfile instructions and enhance build efficiency through instruction reordering.
    \item \added{We evaluated \toolname on 2,000 popular repositories, achieving an average optimization time of 77.55 seconds per Dockerfile, a 26.5\% reduction in rebuild time, and only 0.21\% of optimizations broke compatibility.} 
    \item We identified and categorized four recurring optimization patterns in Dockerfiles, which can serve as practical guidelines for Dockerfile development and a foundation for future studies.
    \item We have open-sourced our dataset and tools ~\cite{home-page}, representing the first dataset on Dockerfile dependencies and modification frequency, to facilitate further research by the community.
\end{itemize}


\section{Background And Motivation}
\label{sec:motivation}

\subsection{Terminology}

\noindent $\bullet$ \textbf{\added{Dockerfile and Instructions}}. \added{A Dockerfile contains all the instructions that a user can invoke on the command line to assemble a Docker image~\cite{dockerfile-official-document}. By using \texttt{docker build}, users can create an automated build that executes a series of command-line instructions to construct a Docker image. These Dockerfile instructions define the steps for building the image~\cite{dockerfile-instruction-document}. Each instruction follows a standardized syntax, beginning with a keyword, followed by arguments that specify the action or configuration. Based on their functionality~\cite{dockerfile-official-document}, Dockerfile instructions can be categorized into eight types: configuration \deleted{(e.g., \texttt{FROM}, \texttt{ARG}, \texttt{ENV}, \texttt{LABEL})}, file-system management \deleted{(e.g., \texttt{COPY}, \texttt{ADD}, \texttt{WORKDIR}, \texttt{USER}, \texttt{VOLUME})}, execution/lifecycle \deleted{(e.g., \texttt{RUN}, \texttt{SHELL}, \texttt{CMD}, \texttt{ENTRYPOINT})}, and networking/health. \deleted{(e.g., \texttt{EXPOSE}, \texttt{ONBUILD}, \texttt{HEALTHCHECK}, \texttt{STOPSIGNAL})}}

\noindent $\bullet$ \textbf{\added{Docker Image and Build Cache}}. \added{A Docker image is a lightweight, standalone, and immutable file that includes the executable application and its required environment, such as system libraries and tools, to run the application~\cite{docker-image-official-document, zhu2024docseckg}. Images are used to create Docker containers and are generated through the process of building a Dockerfile. The Docker build cache accelerates the image-building process by preserving intermediate layers from previous builds~\cite{docker-build-cache-official-document}. When a build command is executed, Docker reuses unchanged cached layers, reducing build time by only rebuilding modified layers, thus improving build efficiency.}

\subsection{Motivating Example}

\begin{figure}[t]
    \centering
    \includegraphics[width=\columnwidth]{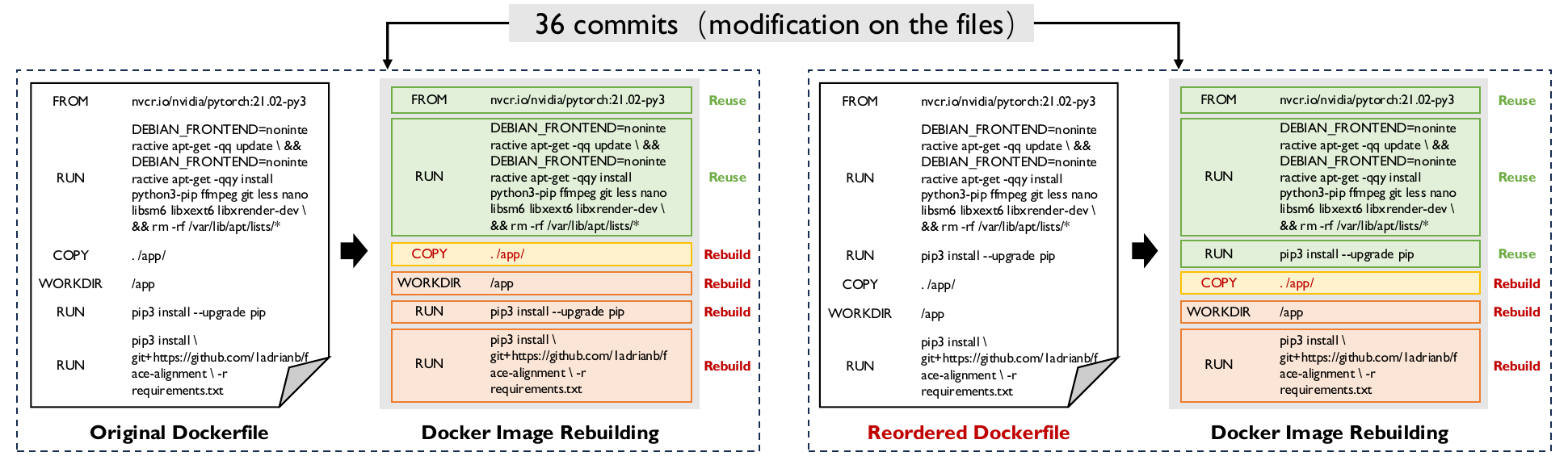}
    \caption{Motivating Example}
    \label{fig:motivation}
    \Description{Motivating Example}
\end{figure}

\added{The Docker image build process relies on a caching mechanism that reuses previously built layers unless modifications occur. This caching speeds up subsequent builds by rebuilding only the layers affected by changes, making the instruction order in the Dockerfile critical to rebuild costs. For example, the case in ~\Cref{fig:motivation} illustrates a Dockerfile from a GitHub repository ~\cite{motivating-example} with 36 commits since its last modification. Modifications to files can trigger a rebuild of the \texttt{COPY} layer, causing all subsequent layers to be rebuilt.} \replaced{By postponing frequently modified instructions and prioritizing stable, resource-intensive steps, unnecessary cache invalidations are minimized.}{To optimize the Dockerfile sequence, instructions can be reordered based on their modification frequency and build cost. By delaying frequently modified instructions and advancing resource-intensive yet stable steps, unnecessary cache invalidations are minimized.} In the right case in ~\Cref{fig:motivation}, the \texttt{RUN pip3 install} instruction has been moved forward, which reduces the rebuild costs by limiting the number of rebuilt layers.

\subsection{Problem Definition}

Consider a Dockerfile sequence \( S \) consisting of a set of \( n \) commands, where \( S = \{c_1, c_2, ..., c_n\} \). Each command \( c_i \) in the sequence is attributed with a modification frequency \( f_i \), reflecting the likelihood of alteration, and a modification cost \( b_i \), indicating the resource consumption when the command is independently executed. The inter-command dependencies are dictated by a set of partial order relations \( \mathcal{R} \), where \( \mathcal{R} \subseteq \{ (c_i, c_j) \mid c_i, c_j \in S \text{ and } i \neq j \} \), such that \( c_i \prec c_j \) for each \( (c_i, c_j) \in \mathcal{R} \), signifying that command \( c_i \) must be executed before command \( c_j \). The aggregate modification cost \( T_i \) for command \( c_i \) is delineated as:

\begin{equation}
    T_i = f_i \times \sum_{k=i}^{n} b_k
\end{equation}

The cumulative cost \( C(S) \) for the sequence \( S \) is thus defined as the summation of aggregate modification costs across all commands:

\begin{equation}
    C(S) = \sum_{i=1}^{n} T_i
\end{equation}

The objective of the problem is to ascertain a new command sequence \( S' \), which is a permutation of \( S \) that honors the partial order constraints in \( \mathcal{R} \) and diminishes the total cost \( C(S') \) to a minimum. \added{This cumulative cost accurately reflects rebuild costs when multiple changes occur. The model prevents double-counting by attributing the rebuild cost of an instruction only once, based on the earliest instruction that triggers the rebuild. When multiple instructions are modified simultaneously, the rebuild cost is effectively captured by the first modified instruction, which triggers the rebuild of subsequent instructions. Any subsequent changes are inherently accounted for in the cost of the earlier modification, ensuring that the total rebuild cost is not overestimated.}

\section{Methodology}
\label{sec:methodology}

\begin{figure*}[t]
    \centering
    \includegraphics[width=0.95\textwidth]{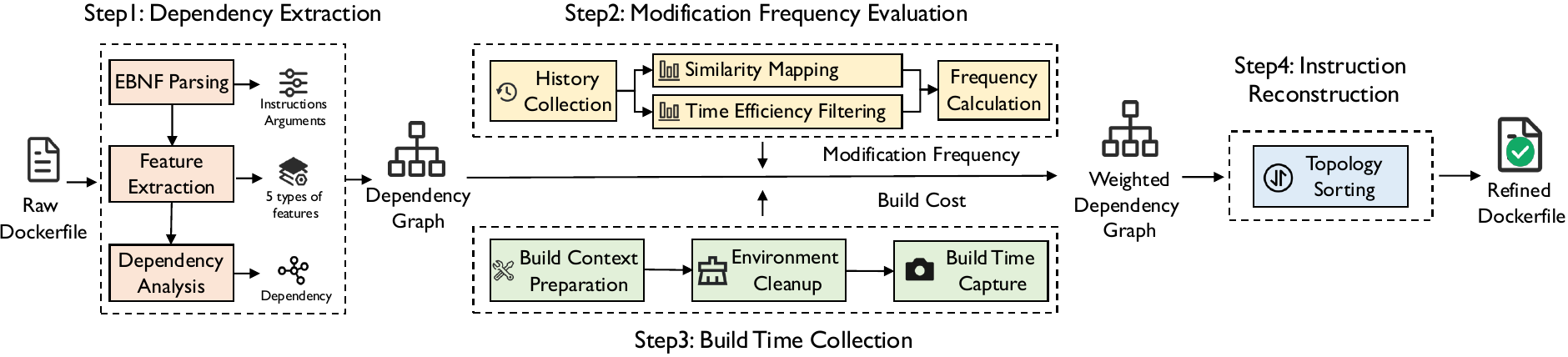}
    \caption{Overviews of \toolname}
    \label{fig:overviews}
    \Description{Overviews of \toolname}
\end{figure*}

\Cref{fig:overviews} presents the overviews of the \toolname, which contains four steps. Firstly, we analyzed the dependency between instructions. By parsing the grammar of the Dockerfile and the inner shell script, we extracted four related features of each instruction, including variables, users, paths/files, and context. Based on these features, we determined the dependency between every two instructions. Subsequently, we evaluate each instruction's modification frequency based on the version control system. We adopt a quantitative approach, assessing the likelihood of future modifications by considering both similarity to past changes and the time efficiency, encapsulated through specific metrics. Then, we collected the build time of each instruction. \replaced{To ensure accuracy, we performed a thorough environment cleanup and conducted multiple trials, yielding precise and reliable data.}{To ensure accuracy in this measurement, we undertake a rigorous cleanup of the built environment and conduct repeated trials, thus obtaining precise and reliable build time data.} Finally, we aligned the modification frequency and build cost for a weighted dependency graph. We designed a topological sorting algorithm to reconstruct the dependency graph into a sequence, which can achieve the lowest build cost in total while obeying the constraints.

\subsection{Dependency Extraction}

In this section, we meticulously extracted the intrinsic dependencies between Dockerfile instructions. Initially, following the official grammar definition, we parsed the raw instructions by using Extended Backus–Naur Form (EBNF) ~\cite{EBNF}. Then, we extracted the elements and summarized them into five types, which can be used to determine the dependency. Next, based on the extracted elements, we proposed a taxonomy of Dockerfile instruction dependency. Finally, we mapped the instructions with their potential dependencies and designed related rules to judge each instruction pair, which constructed the final dependency graph.


\subsubsection{Grammar Parsing}
Accurate parsing of the Dockerfile syntax is the first step for dependency analysis. Dockerfiles are written in a Domain-Specific Language (DSL) combined with SHELL scripts ~\cite{dockerfile-reference}, following the general format: \texttt{<INSTRUCTION> [--flag] <arguments>}. Given the complexity of the arguments, which may include strings, key-value pairs, or intricate shell scripts, parsing Dockerfile is a challenging task since there is no official formal representation of Dockerfile grammar ~\cite{dockernogrammar}. To address this, we consulted official documentation and developed a formal grammar model using Extended Backus–Naur Form (EBNF) ~\cite{EBNF}, an enhancement of the Backus–Naur Form (BNF) ~\cite{BNF}. This meta-syntax notation aids in describing the context-free grammar of formal languages. We delineated the expected formats for each instruction, along with their potential flags and arguments. Here is an example of the \texttt{ENV} instruction's grammar. 

\noindent\fbox{
 \footnotesize
 \parbox{0.95\linewidth}{
  \texttt{
    ENV\_INSTRUCTION = "ENV", space, (key, space, value | \{ key\_value\_pair, space \}); \\
    key\_value\_pair = key, space, value;  key = string;  value = string;\\
    string = "'", \{ character \}, "'" | '"', \{ character \}, '"' | \{ character \}; \\
    character = ? any character except newline and unescaped quote ?;  space = " " , \{ " " \};  
    } 
  }
}

Besides basic grammar, shell scripts are also common in Dockerfiles, which offer considerable flexibility and require handling. Defining a grammar model for them is intricate and prone to errors. To overcome this, we utilized third-party tools, libdash ~\cite{libdash}, to parse shell commands into AST. We adopted the labels of each node and recorded the relative information (i.e., command, flags, and arguments). For those complex shell commands that involve control flows (i.e., command sequences connected by \texttt{\&\&} or the pipe symbol \texttt{|}), we segmented them into distinct commands. 
\replaced{By separately processing the DSL and shell scripts, we efficiently parsed Dockerfile syntax and categorized each token, which forms a foundational step for subsequent feature extraction.}{By separately processing the DSL and shell scripts, we efficiently parsed Dockerfile syntax and categorized each token. This forms a foundational step for subsequent feature extraction.}

\subsubsection{Semantic Elements Extraction}
Instructions contain rich semantic information, which can be used to determine the dependencies between instructions. According to the runtime actions and the specified static environment defined by the instructions, we obtained the semantic elements and divided them into five categories. 

    \noindent $\bullet$ 
        \textbf{Variables Sets}: Identifying variables present in an instruction and categorizing them into \texttt{definition} and \texttt{use-only} types. \texttt{Definition} indicates the instruction creates a new variable, while \texttt{use-only} signifies the instruction utilizes a previously defined variable.
        
    \noindent $\bullet$
        \textbf{Related Path or Files}: Determining the absolute paths of files or directories referenced in the instruction. These are further classified into input and output paths based on their creation relationship. The context directory, potentially altered by \texttt{WORKDIR}, is also noted.
        
    \noindent $\bullet$
        \textbf{User}: Identifying the executor of the instruction, with the default user \texttt{root}. Changes in the executing user, such as through the \texttt{USER} instruction or user-related shell commands are recorded.
        
    \noindent $\bullet$
        \textbf{Packages, Libraries, and Tools}: Logging the packages, libraries, and tools used or installed by an instruction, which are divided into \texttt{install} and \texttt{use-only}. For example, \texttt{apt install wget} in one \texttt{install} instruction followed by \texttt{wget https://example.com} (i.e., a use-only instruction).
        
    \noindent $\bullet$
        \textbf{Context Information}: Record the context information of the current layer.

Based on the token parsed in the previous step, we carefully extract the above semantic elements for subsequent dependency judgment. While the above elements are directly extracted from the Dockerfile, certain implicit dependencies require additional knowledge. To address these nuances, we supplied additional semantics, followed by these steps:

\noindent \textbf{Environment Initialization}.
Environment variables defined by \texttt{ARG} and \texttt{ENV} instructions are handled by substituting them with their assigned values. All parameters are parsed into key-value pairs and incorporated into a global dictionary, which serves as a reference to replace variables across other instructions. This approach ensures that elements are not omitted in instructions that rely on environment variables. For instance, the \texttt{RUN} instruction might depend on a path such as "/home/python/1.0.0/", which cannot be directly extracted from the unprocessed instruction.

\noindent\fbox{
 \footnotesize
 \parbox{0.95\linewidth}{
  \texttt{
    ARG VERSION 1.0.0 | ENV HOME\_DIR "/home/python/\$\{VERSION\}/" | RUN cd \$\{HOME\_DIR\}
    } 
  }
}

\noindent \textbf{File Path Expansion}.
We process relative file paths specified in parameters, converting them to absolute paths based on the current working directory. This step ensures that all file references are unambiguous and accurately represented. For \texttt{ADD} and \texttt{COPY} instructions, the source files or directories are expanded to their absolute paths, with wildcard characters considered, and based on this, a dictionary tree structure is constructed to represent the file mappings.

\noindent \textbf{Shell Command Parsing}.
\texttt{RUN} instructions frequently contain complex shell commands that require parsing to identify underlying actions, such as software installations, file manipulations, or configuration adjustments. To achieve this, we decompose commands into their component parts, capturing a comprehensive set of semantic elements for each action. This process employs a third-party library (e.g., libdash ~\cite{libdash}) to interpret shell syntax and extract commands, flags, options, and file manipulations. For frequently used commands, sub-commands are gathered from "man pages" ~\cite{man-page} to retrieve secondary information on packages, libraries, or files.

\subsubsection{Dependency Determination Rules Design}


\begin{table}[t]
    \centering
    \caption{Potential dependencies of each instruction}
    \label{tab:potential-dependency}
    \footnotesize
    \scalebox{0.8}{
    \begin{tabular}{p{4cm}cccccc}
        \toprule
            Instruction          & Variable   & P\&F        &  User      & P\&L\&T                    & Context    &  Other     \\ \midrule
            ARG                  & \checkmark &             &            &                            &            &            \\ 
            ENV                  & \checkmark &             &            &                            & \checkmark &            \\ 
            COPY / ADD / VOLUME  &            & \checkmark  &            &                            &            &            \\ 
            USER                 &            &             & \checkmark &                            &            &            \\ 
            RUN                  & \checkmark & \checkmark  & \checkmark &  \checkmark                & \checkmark &            \\ 
            WORKDIR / EXPOSE     &            &             &            &                            & \checkmark &            \\ 
            HEALTHCHECK          &            & \checkmark  &            &                            &            & \checkmark \\ 
            EXPOSE               &            &             &            &                            & \checkmark &            \\ 
            CMD / ENTRYPOINT     & \checkmark &             &            &                            & \checkmark &            \\ 
            SHELL                & \checkmark & \checkmark  & \checkmark &  \checkmark                & \checkmark &            \\ 
            FROM / ONBUILD / STOPSIGNAL &     &             &            &                            &            & \checkmark \\ 
            LABEL/MAINTAINER     &            &             &            &                            &            &            \\
        \bottomrule
    \end{tabular}}
    \begin{enumerate}
        \item[*] P\&F: Paths and Files. P\&L\&T: Packages, Libraries, and Tools.
    \end{enumerate}
\end{table}

According to the above semantic element classification, we further design rules for each corresponding dependency determination: 

    \noindent $\bullet$ 
        \textbf{Variable-based Dependency}: Determine based on the type of the extracted semantic element. For the same variable, all use-only types elements depend on the definition-type element.
    
    \noindent $\bullet$
        \textbf{File/Directory-based Dependency}: Determine based on the input/output type of the extracted element. For the same file, all input-type elements depend on output-type elements. For paths, the parent path contains the child path.

    \noindent $\bullet$
        \textbf{User-based Dependency}: When processing user-related instructions, the user is stored as a global variable, which defaults to root. When an instruction is modified, this variable is modified, and the user of all subsequent instructions is set to the new value.

    \noindent $\bullet$
        \textbf{Package Manager-based Dependency}: Similar to the rules for handling files, for the same package/library/tool, all use-only type elements depend on the installation type elements.
    
    \noindent $\bullet$
        \textbf{Context-based Dependency}: Like processing User, all global semantic variables are stored. When a modification occurs, all subsequent instructions depend on the modification instruction.

    \noindent $\bullet$
        \textbf{Other Dependency}: These are dependencies related to specific conditions or instructions not fitting into the above categories. They often involve fundamental requirements or global dependencies in Docker builds, including \texttt{FROM}, \texttt{HEALTHCHECK}, and \texttt{ONBUILD} dependency.

Based on the dependency classification above, we further analyzed the potential dependencies of each instruction, as shown in ~\Cref{tab:potential-dependency}.

\subsubsection{Dependency Analysis}
Following the summarization of the potential dependency of each instruction, we formulated rules to ascertain the presence of dependencies between any two instructions. Considering that Dockerfile encompasses 18 types of instructions (except \texttt{MAINTAINER}, which is deprecated ~\cite{dockerfile-reference}), this yields 324 potential instruction pair combinations. Notably, not every combination implies a dependency, such as the \texttt{EXPOSE} instruction, which solely signifies port openings and does not depend on other instructions.

To streamline the analysis, we employed a two-step approach. Initially, we focused on the types of instructions. This preliminary step efficiently filters out combinations that invariably possess dependencies (like \texttt{FROM-RUN}) and those that unequivocally do not (such as \texttt{LABEL-RUN}), circumventing the need for an in-depth dependency analysis. Subsequently, for the remaining combinations, we matched the corresponding semantic elements against the predefined potential dependency types of the instructions, as outlined in ~\Cref{tab:potential-dependency}. 


\subsubsection{Implementation Example}
\begin{figure*}[t]
    \centering
    \includegraphics[width=0.9\textwidth]{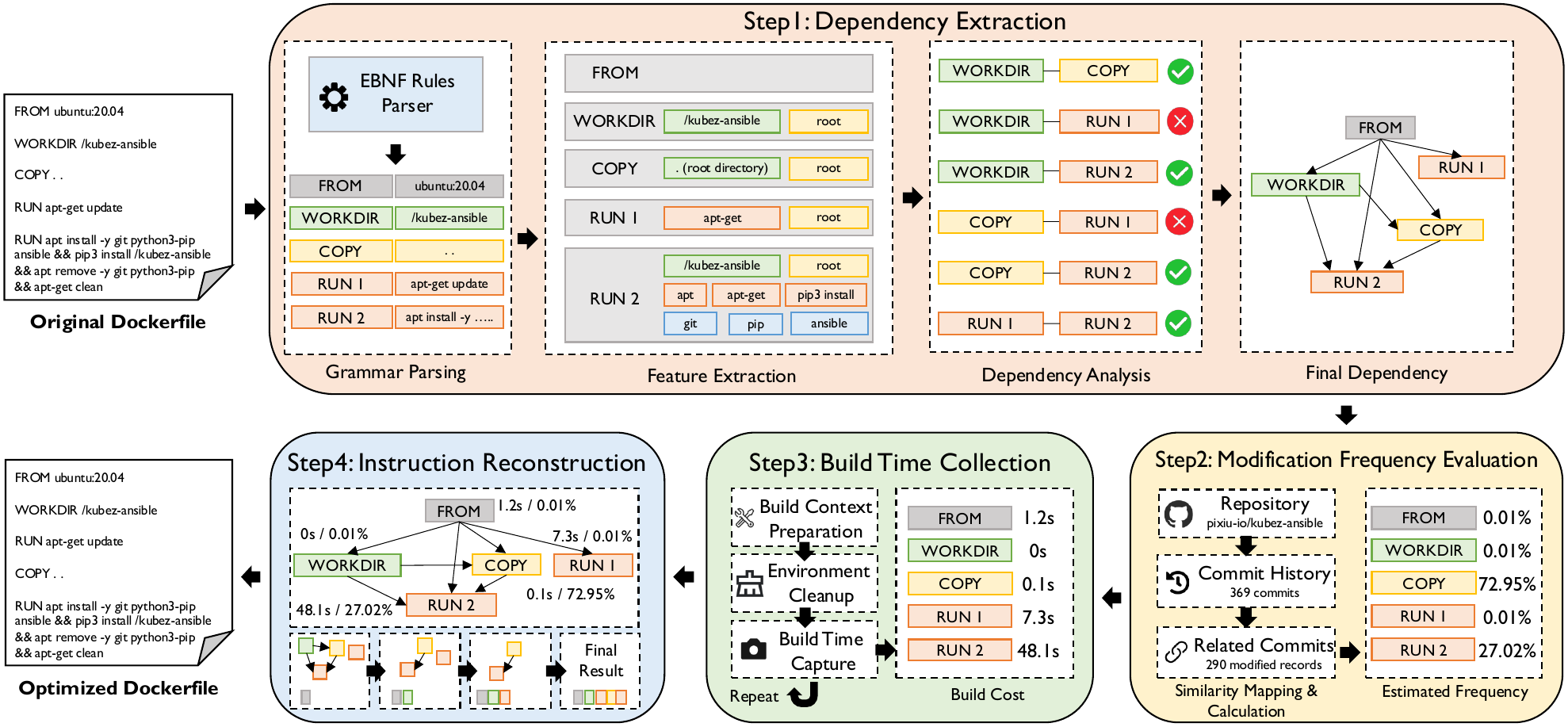}
    \caption{Optimization Example of \toolname}
    \label{fig:example}
    \Description{Optimization Example of \toolname}
\end{figure*}

As shown in ~\Cref{fig:example}, for a given Dockerfile~\cite{kubez-ansible}
, we firstly parsed the original Dockerfile through EBNF rules parser. Then, we extracted each feature type, which is marked by different colors. Finally, based on the feature information and the potential dependencies of each instruction, we analyzed each instruction pair and constructed the dependency graph.

\subsection{Modification Frequency Evaluation}
In this section, we recognize that modifications to Dockerfile instructions precipitate the rebuilding of subsequent instructions. \replaced{Placing instructions with high modification frequency at the beginning of a Dockerfile can lead to unnecessary and repetitive rebuilds. Although future modifications cannot be predicted with certainty, historical modification data provides a valuable proxy for forecasting potential changes.}{Placing an instruction with a high frequency of changes near the start of a Dockerfile can result in unnecessary and repetitive rebuilding processes. While it is impossible to predict future modifications with absolute certainty, historical modification data serves as a valuable proxy for forecasting potential changes.} To capitalize on this, we meticulously gathered modification records from version control systems. Employing methodologies that consider both the similarity of past modifications and their temporal relevance, we quantitatively assessed the modification frequency for each instruction in the current version of the Dockerfile.

\subsubsection{Modification History Collection}
\deleted{In our analysis of Dockerfiles hosted on GitHub, the initial step involved cloning the respective repositories to local storage. Subsequently, we collected MD5 hashes for all commit histories, utilizing the \texttt{git diff} command to ascertain the modifications occurring between adjacent commits. Given that Dockerfiles are predominantly named \texttt{Dockerfile}, we streamlined our process by filtering modification records related explicitly to Dockerfiles via keyword mapping. Each modification record was meticulously documented, capturing details such as the commit ID, instruction type, content of the instruction, date of modification, nature of the modification (e.g., addition, deletion, modification), and the respective line numbers.}

\deleted{Furthermore, we acknowledged and addressed the presence of implicit modifications pertinent to Dockerfiles. Certain instructions within Dockerfiles, such as \texttt{ADD} and \texttt{COPY}, necessitate interactions with the local file system, thereby rendering modifications to the corresponding files or directories as triggers for instruction rebuilding. To comprehensively track these modifications, we established and maintained a list encompassing all addresses, both direct and indirect, involved in the Dockerfiles. Modifications to any files or directories enlisted were then duly recorded as alterations associated with the corresponding Dockerfile instruction.}

\added{In analyzing Dockerfiles hosted on GitHub, the initial step involved cloning the respective repositories to local storage. We then collected MD5 hashes for all commit histories and used the \texttt{git diff} command to identify modifications between adjacent commits. Since Dockerfiles are typically named \texttt{Dockerfile}, we streamlined the process by filtering modification records related specifically to Dockerfiles through keyword mapping. Each modification record was meticulously documented, capturing details such as the commit ID, instruction type, content, modification date, \replaced{type}{nature} of the change (e.g., addition, deletion, modification), and the relevant line numbers. Additionally, we recognized and addressed implicit modifications related to Dockerfiles. Certain instructions, such as \texttt{ADD} and \texttt{COPY}, interact with the local file system, making changes to associated files or directories triggers for instruction rebuilding. To comprehensively track these modifications, we maintained a list of all addresses, both direct and indirect, referenced within the Dockerfiles. Any changes to files or directories listed were recorded as modifications linked to the corresponding Dockerfile instruction.}

\subsubsection{Modification Frequency Estimation}

To assess the modification likelihood of Dockerfile instructions, we analyzed historical modifications that share characteristics similar to those of the current version’s instructions. Our hypothesis suggests that the frequency of similar historical modifications provides a reliable indicator of the probability of future changes to the current instruction. To quantify the probability, we employed textual similarity as the metric to evaluate the relationship between each current instruction and its historical counterparts.

However, textual similarity alone may not accurately capture the semantic differences between some instructions, where minor textual variations can result in significant semantic shifts. To address this limitation, we introduce a classification-based approach that categorizes Dockerfile instructions into four distinct types, each with specific comparison rules:

\begin{itemize}[leftmargin=*]
    \item 
        \textbf{Key-Value Pair Instructions:} These follow a key-value format where the key must strictly match, as any change to the key alters the instruction’s semantics. For these instructions, we enforce exact matching of the keys, disregarding variations in the values (i.e., \texttt{ARG, ENV, USER, EXPOSE, LABEL}).
    \item 
        \textbf{File-System Instructions:} These involve file system operations, where semantic relationships depend on file path containment. A parent path can match a child path, but a child path cannot match a parent path, ensuring accurate capture of file-level modifications. For these instructions, we only consider the matching records (i.e., \texttt{COPY, ADD, VOLUME, WORKDIR, ENTRYPOINT}).
    \item 
        \textbf{Shell-Script Instructions:} Simple text similarity is often insufficient for these instructions, as minor changes can have substantial semantic effects. We apply threshold-based filtering to differentiate between minor and significant modifications (i.e., \texttt{RUN, SHELL, CMD}).
    \item 
        \textbf{Special Instructions:} These instructions, with any change, typically have a significant impact on the Dockerfile’s behavior. Therefore, we handle these instructions with direct type-based matching (i.e., \texttt{FROM, ONBUILD, HEALTHCHECK, STOPSIGNAL}).
\end{itemize}


For the strict matching instructions, the similarity is considered as 1. For the rest of instruction \( c \), we computed the textual similarity, \( \text{Sim}(c, c') \), between the instruction \( c \) and each historical modification record \( c' \) of the same type. The process involves vectorizing each instruction using TF-IDF (Term Frequency-Inverse Document Frequency) and subsequently calculating the cosine similarity between the vectors of the instruction \( c \) and the modification record \( c' \). 

\subsubsection{Time Efficiency Filtering}
The relevance of data in the context of temporal dynamics is a critical factor in our analysis, particularly when considering the modification history of Dockerfile instructions. Drawing from established principles in data mining and recommendation systems, it is understood that recent data tends to be more pertinent to current queries, while the significance of older data diminishes over time ~\cite{decay, decay-app1, decay-app2, decay-app3}. In alignment with this premise, we posit that recent modifications in Dockerfile instructions are more indicative of potential future changes compared to modifications that transpired in the distant past. Since then, we have only considered the record within \replaced{the past 30 months, which is experimentally quantified in ~\Cref{rq:1.1}.}{three years}. 

\subsubsection{Modification Frequency Calculation}
To quantify the modification propensity of a specific Dockerfile instruction \( c \), we calculate its overall modification frequency \( F(c) \). This computation aggregates the weighted similarities of historical modifications, further normalized by the total number of Dockerfile-related changes. The formula for this calculation is articulated as:

\begin{equation}
    F(c) = \frac{\sum_{c'} \text{Sim}(c, c')}{\text{Total Modifications}}
\end{equation}

Post-calculation of the modification frequency \( F(c) \) for each instruction, we embark on a normalization process. This step is imperative for elucidating the relative modification likelihood of each instruction within the entire Dockerfile context. Normalization is executed by dividing the modification frequency of an individual instruction by the aggregate of modification frequencies across all instructions within the Dockerfile. Consequently, this yields a normalized modification frequency \( F_{\text{norm}}(c) \) for each instruction, defined by:

\begin{equation}
    F_{\text{norm}}(c) = \frac{F(c)}{\sum_{c \in \text{Dockerfile}} F(c)}
\end{equation}

In this equation, \( \sum_{c \in \text{Dockerfile}} F(c) \) represents the cumulative sum of modification frequencies for all Dockerfile instructions. This normalization process scales each instruction’s frequency to a value between 0 and 1, facilitating a comprehensible, normalized gauge of its modification likelihood relative to other instructions. 

\subsubsection{Implementation Example}
Continue the example in ~\Cref{fig:example}. Firstly, we cloned the original repository and got all the commit records. Through all 369 commits, we filtered out 290 related to the Dockerfile, including the direct modification of the Dockerfile and the modification of the mentioned files. Then, for each related commit, we calculated the similarity to the current instruction and summed it up for each instruction. Finally, we performed normalization calculations to obtain the final modification frequency of each instruction.


\subsection{Build Time Collection}

The precise measurement of build times for each Dockerfile instruction is pivotal in our Dockerfile reconstruction analysis. In this section, we systematically construct Dockerfile images, capturing the build time associated with each instruction. Recognizing the potential influence of various environmental factors on these build times, we implement a rigorous protocol for environment cleanup before each build. This ensures that each build process starts from a standardized baseline, free from any residual cache or data that might skew the results. Furthermore, to account for and neutralize the impact of inherent variances in the build process, we adopt a strategy of repeated builds. This approach allows us to calculate an average build time for each instruction, thereby yielding more reliable and consistent data.

\subsubsection{Build Environment Cleanup}
Before each Docker build, we meticulously ensured a clean build environment to negate the influences of existing build caches and base images, which are known to impact build times significantly. This process commenced with the deletion of all extant Docker images and containers. Subsequently, we utilized the \texttt{docker system prune -a} command, a functionality provided by the Docker Command Line Interface (CLI), to comprehensively remove all unused Docker entities, including containers, networks, images, and volumes ~\cite{docker-prune}. The cleanup process culminated with executing the \texttt{docker system df} command ~\cite{docker-system-df}, serving as a verification step to confirm the complete eradication of any residual build cache. This meticulous cleanup protocol ensures a standardized baseline for each build, thereby enabling accurate and consistent measurement of build times.

\subsubsection{Build Time Capture}
With the build environment's effect neutralized, our focus shifted to the construction of the target Docker image and the precise measurement of the time consumed for each layer's build. The build process inherently generates logs, outputted to the console, which became our primary data source for capturing the build times. With the advent of Docker Engine version 23.0, BuildKit emerged as the default builder, introducing a segmented build process delineated into distinct stages, each marked by sequential order. Completion of each stage is signified in the console output with a message following the format: \texttt{\#[number] DONE [time]s}. In this context, "number" represents the order of the completed stage, and "time" indicates the duration in seconds. Leveraging regular expressions, we matched these log entries and correlated each stage's order with its corresponding Dockerfile instruction.

Acknowledging that external factors such as network conditions can also influence build times ~\cite{wu2022understanding}, we implemented a procedure to mitigate these variances. This entailed repeating the build process for each image three times and subsequently computing the average build time for each instruction. Through this approach, we acquired a reliable measure of the specific build time for each Dockerfile instruction, thereby enhancing the precision of our build time analysis.

\subsubsection{Implementation Example}
Keep on the example in ~\Cref{fig:example}. The build context was already settled in the previous steps. We first cleaned up the runtime environment and emptied all the caches and images. Then, we initialized the Dockerfile into the Docker image, capturing the time cost for each instruction. After three repeating trails, we got the average build cost for each instruction.


\subsection{Instruction Reconstruction}
After obtaining the modification frequency and predictive build cost for each instruction, we aligned them to the related nodes and got a weighted dependency graph. Based on this, we designed a topological sorting algorithm to serialize it. The algorithm initializes by constructing a weighted directed graph from the declared dependencies. A priority queue is then employed to facilitate the dynamic selection of nodes based on their calculated cost, a product of the modification frequency, and the aggregated build times of the remaining commands.

\begin{algorithm}[t]
\caption{Topological Sorting-Based Optimization of Dockerfile Instructions}
\label{alg:topological_sort_optimization}
\footnotesize

\begin{flushleft}
\textbf{Input}: Directed Weighted Graph $G(V, E)$, priority queue $Q$, $indegree$ \\
\textbf{Output}: Optimized sequence of instructions
\end{flushleft}

\begin{algorithmic}[1]
\WHILE{$Q \neq \emptyset$}
    \STATE $v \gets Q.\text{pop\_min}()$
    \STATE $optimized\_sequence.\text{add}(v)$
    \FOR{each neighbor $w \in G[v]$}
        \STATE $indegree[w] \gets indegree[w] - 1$
        \IF{$indegree[w] = 0$}
            \STATE $cost[w] \gets frequency[w] \times \sum\limits_{u \in r\_n} build\_time[u]$
            \STATE $Q.\text{insert}(w, cost[w])$
        \ENDIF
    \ENDFOR
    \STATE $remaining\_nodes.\text{remove}(v)$
\ENDWHILE

\RETURN $optimized\_sequence$

\end{algorithmic}
\end{algorithm}

Initially, the indegree of each node is computed to identify nodes with no dependencies, which are then added to the priority queue. The cost for each node is calculated, taking into account the frequencies and build times, thereby prioritizing nodes with lower costs for early execution. This step ensures that the execution order respects the dependency graph while also optimizing the build process. As the algorithm progresses, nodes are dequeued and added to the optimized sequence. When a node is dequeued, it signifies the completion of its corresponding instruction, prompting a recalculation of the costs for its dependent nodes. The recalculated costs reflect updated build times as the remaining instructions in the graph are processed. Nodes with updated indegrees of zero are then re-evaluated for their costs and added to the queue. This iterative process continues until the queue is empty.

\added{It is important to note that, in practical scenarios, developers often group semantically related instructions together, creating "instruction groups" for easier understanding and maintenance. For instance, when setting up a Java environment, instructions such as updating the environment (e.g., \texttt{apt update}), downloading the package (e.g., \texttt{apt install jdk}), and defining environment variables (e.g., \texttt{JAVA\_HOME}) are typically written together. However, their modification frequencies can vary significantly, causing the original "instruction group" to be disrupted after reordering, which decreases code readability. To mitigate this, we propose enhancing the semantic readability of a Dockerfile by incorporating its extracted dependency tree. The aforementioned semantic groups can be represented as subtrees within the dependency tree, thereby compensating for the readability loss in the flattened sequence.}

The result of this algorithm is an optimized sequence of Dockerfile instructions that not only adheres to the necessary dependency constraints but also minimizes the overall build cost. This approach leverages the efficiencies of topological sorting and cost-based prioritization, making it particularly effective for optimizing Dockerfile sequences in scenarios with complex dependencies and varying instruction costs.

\subsubsection{Implementation Example}
Still, take the example in ~\Cref{fig:example}. After the previous steps, we attached the build cost and modification frequency to the dependency graph. Based on the weighted dependency graph, we implemented the topology sorting and got the final optimized sequence.

\section{Evaluation}
\label{sec:evaluation}

In this section, We evaluate the effectiveness and performance of \toolname by answering the following research questions (RQs).

   \noindent $\bullet$ \textbf{RQ1: Effectiveness \& Efficiency}. \deleted{How does \toolname perform on the Dockerfile reconstruction in practice?}\added{How does \toolname perform compared to existing tools on repositories of varying quality?} \added{And, how is \toolname's operational efficiency and usage frequency in practical applications?}
   
    \noindent $\bullet$ \textbf{RQ2: Consistency Analysis.} Does the Dockerfile maintain functional equivalence after optimization?
   
    \noindent $\bullet$ \textbf{RQ3: Ablation Study.} How does each step affect the optimization \deleted{efficiency} and build success rate?
    
    \noindent $\bullet$ \textbf{RQ4: Contribution Analysis.} What prevalent patterns contribute most to the optimization\deleted{of Dockerfile}?

\deleted{RQ1 investigates the overall effectiveness and performance of \toolname in real-world cases, RQ2 inspects the behavior equivalence before and after Dockerfile optimization by \toolname, RQ3 conducts an ablation study to investigate which steps in our methodology signifies the overall improvement in performance and behavior consistency assurance of optimization, and RQ4 identifies the major optimization patterns that contribute the most to the optimization.}


\noindent \textbf{Experiment Data Set Preparation.} We select GitHub as the data source to construct the experiment data set. \deleted{We set a filter with a stargazer count greater than 1000 to filter out repositories that are widely popular with the community.} \added{Considering that repository quality may have a certain impact on the optimization results, we selected repositories of different quality based on the number of stars. Specifically, we randomly selected 500, 500, and 1000 repositories containing Dockerfiles from three ranges: 0-500 stars, 500-1000 stars, and over 1000 stars, respectively.} \deleted{Further, we filter out the repositories that do not adopt Docker in deployment by whether they include Dockerfile.} Since confirming build context in multiple Dockerfiles within the same repository ~\cite{henkel2021shipwright}, we only select repositories whose Dockerfile is in the root directory, followed by the previous study ~\cite{wu2022understanding}. \deleted{After these criteria, 3,781 repositories were left.} Further, we cloned all the repositories locally and built the latest version of the Dockerfile. \deleted{We randomly selected 1,000 repositories from the successful case to build the experiment data set.} 

\noindent \textbf{\added{Comparison Tools.}} \added{Existing work on Dockerfile optimization mainly focuses on eliminating code smells ~\cite{ksontini2025refactoring,durieux2023parfum,bui2023dockercleaner,rosa2024not}, without considering the impact of modification frequency on refactoring efficiency. Therefore, no directly comparable tools exist. As a result, we opted to compare our approach with Docker smell remediation tools. The most widely used tool for detecting Dockerfile smells is Hadolint ~\cite{hadolint}; however, it only performs detection and does not offer repair suggestions. Hence, we selected the latest tools, Parfum ~\cite{durieux2023parfum} and DockerCleaner ~\cite{bui2023dockercleaner}. The experimental process and the calculation method for optimization efficiency were consistent with the approach described above, and the tools were configured with their default settings.}

\noindent  \textbf{Experiment Environments.} All of the experiments were conducted on Ubuntu 20.04.6 LTS with 2.50GHz Intel(R) Xeon(R) Gold 6248 CPU and 188GB RAM. The Docker is 24.0.6, build ed223bc.

\subsection{\replaced{RQ1.1: Effectiveness}{RQ1: Effectiveness \& Efficiency}}
\label{rq:1.1}

\noindent \textbf{\added{Evaluation Metrics.}} \added{To assess the effectiveness of \toolname in enhancing Dockerfile build performance, we conducted experiments on a dataset of Dockerfile samples. We define the optimization efficiency for each project as the sum of optimization efficiencies for each modification, divided by the total number of modifications, as shown in the following formula, where M is the total number of modifications for a given project.}

\begin{equation}
    \text{Optimization Efficiency} = \sum_{i=1}^{M} \frac{\text{Build Time}_{\text{before},i} - \text{Build Time}_{\text{after},i}}{\text{Build Time}_{\text{before},i}} \times \frac{1}{M}
\end{equation}

For each Dockerfile, we compiled all modification records from the past three months (or the 10 most recent modifications, if fewer than 10 existed within this period). For each modification, \toolname was applied to reorder instructions based on both build time and modification frequency. We then measured the rebuild time for the optimized and original versions using the subsequent modification as a reference point. This process was repeated three times per modification, and the average rebuild time was computed. The final optimization outcome for each Dockerfile was determined by averaging the results from all modifications.

\begin{figure*}[t]
    \centering
    \begin{minipage}{0.48\textwidth}
        \centering
        \includegraphics[width=\textwidth]{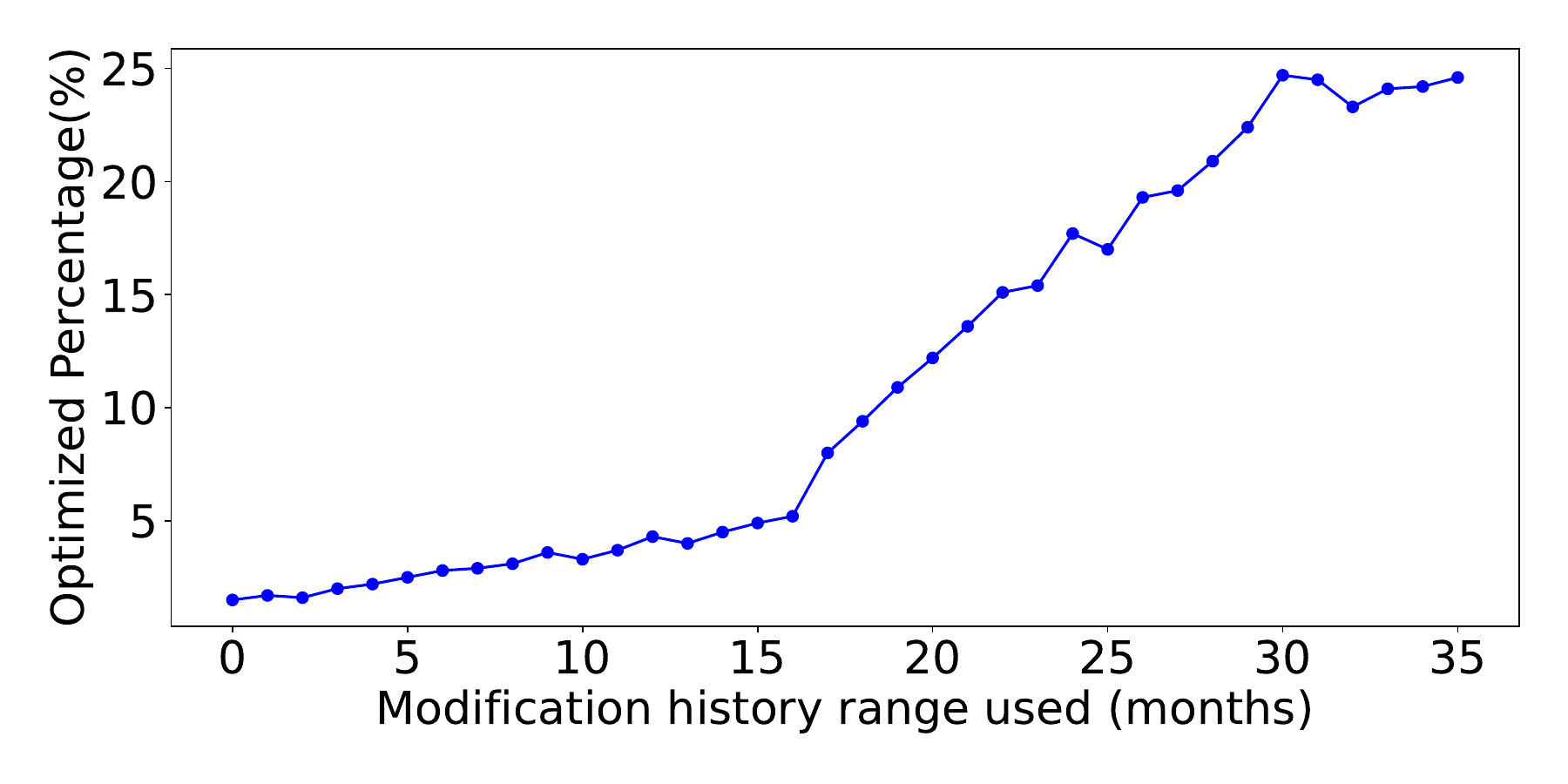}
        \caption{\added{Used Modification Record vs. Optimization}}
        \label{fig:rq1-5}
        \Description{Scope of Used Modification Record vs. Optimization}
    \end{minipage}%
    \hfill
    \begin{minipage}{0.48\textwidth}
        \centering
        \includegraphics[width=\textwidth]{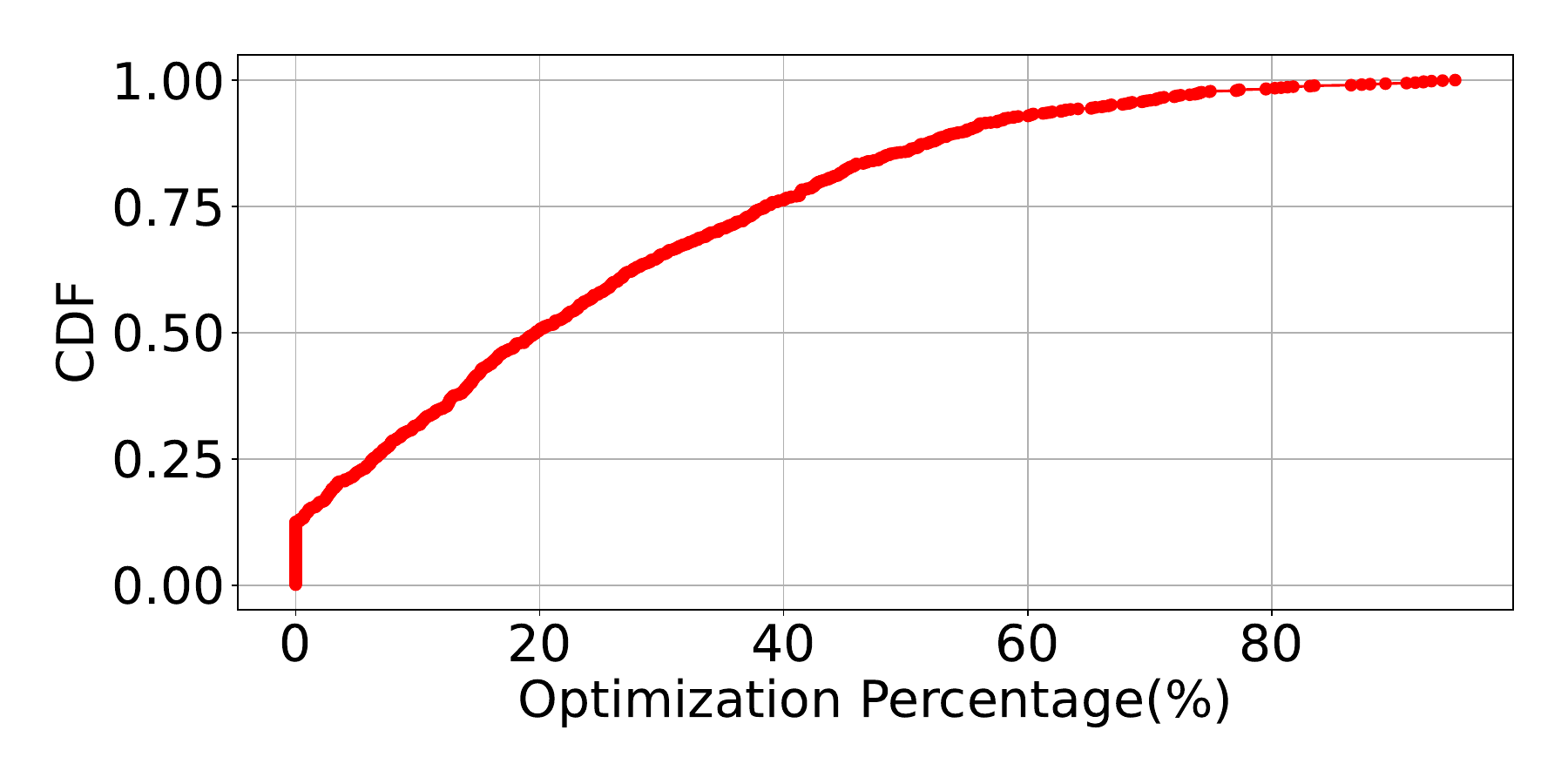}
        \caption{\added{CDF of Optimization Results}}
        \label{fig:rq1-1}
        \Description{CDF of Optimization Results}
    \end{minipage}
\end{figure*}

\noindent \textbf{\added{Record Scope Determination.}} \added{First, we determined the optimal range for modification records through experiments. We incrementally increased the range in monthly units and observed the changes in optimization efficiency. We randomly selected 200 projects from the dataset for testing. The influence on the optimization is shown as ~\Cref{fig:rq1-5}. When the range of modification records used is relatively short (less than 8 months), the optimization effect is less than 5\%. The main reason for this is that the limited amount of data fails to accurately reflect the modification frequencies of the individual instructions, leading to an incorrect calculation of the true loss model during the topological sorting process, and thus suboptimal optimization results. When the record range is approximately 2.5 years (30 months), the tool achieves the optimal effect. Further increasing the data range does not result in a significant improvement in optimization efficiency. Therefore, we conclude that using modification records from the past 2.5 years is the most appropriate.}

\noindent \textbf{\added{Effectiveness Evaluation.}} \added{After determining the scope of the record, we adopted \toolname on our dataset.} During the optimization process, a limited number of rebuilds failed. Specifically, \replaced{27}{13} samples \deleted{(1.3\%)} encountered build failures post-optimization. Manual inspection revealed these failures were due to unavailable external dependencies defined within the image, rather than issues arising from the sequence optimization itself. Excluding these cases, all other Dockerfiles were successfully rebuilt post-optimization without any failures attributable to \toolname's modifications. \added{In all successful cases, \toolname effectively improved 1830 samples (92.75\%) of all successfully built Dockerfiles in the dataset, reducing modification and reconstruction time by an average of 26.5\%. We further analyzed the distribution of the optimization results, as shown in ~\Cref{fig:rq1-1}. The x-axis represents the proportion of build time optimization, and the y-axis represents the probability density. In \replaced{7.25\%}{11.34\%} of the samples (\replaced{143}{112} cases), the reconstruction time remained the same, indicating that these Dockerfiles were already optimally ordered. For \replaced{12.82\%}{14.39\%} of the cases (\replaced{253}{142} samples), the reconstruction time was reduced by more than 50\%, demonstrating significant improvement in build efficiency. }



\noindent \textbf{\added{Performance on Different Record Ranges.}} \added{Furthermore, we explored \toolname's performance across different data ranges. As shown in ~\Cref{tab:performance-ranges}, \toolname achieved the best optimization effect in the 0-500 stars range, with an optimization efficiency of 30.6\%. We manually examined a portion of the samples, and found that the data in the 0-500 stars range had significant room for improvement in Dockerfile quality, which is why the tool performed better in this range. }

\begin{table}[t]
    \centering
    \begin{minipage}{0.48\textwidth}
        \centering
        \caption{\added{Performance across different data ranges}}
        \footnotesize
        \label{tab:performance-ranges}
        \scalebox{0.9}{
            \begin{tabular}{c c c}
                \toprule
                Data Ranges      & Avg. Optimized Percentage & Avg. Time Cost \\ \midrule
                0 - 500 stars    & 30.6\%                    & 62.1s          \\
                501 - 1000 stars & 26.3\%                    & 78.9s          \\
                Over 1000 stars  & 24.5\%                    & 84.6s          \\ 
                \bottomrule
            \end{tabular}
        }
    \end{minipage}
    \hfill
    \begin{minipage}{0.48\textwidth}
        \centering
        \caption{\added{Optimized Percentage Comparison}}
        \footnotesize
        \label{tab:performance-tools}
        \scalebox{0.9}{
            \begin{tabular}{c c}
                \toprule
                Tools & Avg. Optimized Percentage \\ \midrule
                Doctor & 26.5\% \\
                Parfum & 15.3\% \\
                DOCKERCLEANER & 17.9\% \\
                \bottomrule
            \end{tabular}
        }
    \end{minipage}
\end{table}

\noindent \textbf{\added{Existing Tool Comparison.}} \added{As for the comparative experiment with existing tools, as shown in ~\Cref{tab:performance-tools}, the build efficiency of Parfum ~\cite{durieux2023parfum} and DockerCleaner ~\cite{bui2023dockercleaner} improved by 15.3\% and 17.9\%, respectively. We analyzed the reasons behind these improvements. By eliminating code smells, they reduced redundant dependencies or replaced them with more streamlined base images/packages, leading to an average reduction in image size of 11.2\%, which in turn reduced build time. However, due to their lack of consideration for modification frequency, they were unable to effectively utilize the caching mechanism, resulting in limited optimization efficiency.}




\mybox{\textbf{Answer to RQ1.1:}
On average, \toolname increases Dockerfile refactoring efficiency by 26.5\% on 2,000 repositories, outperforming existing code-smell-based tools, which achieve 11.2\% and 8.6\% improvements, respectively. For 12.82\% of the dataset, the reconstruction time was reduced by more than 50\%. Notably, \toolname's effectiveness rises to 30.6\% in lower-quality projects, indicating the relationship between repository quality and optimization effectiveness.}

\subsection{\added{RQ1.2: Efficiency}}

\begin{figure*}[t]
    \centering
    \begin{minipage}{0.48\textwidth}
        \centering
        \includegraphics[width=\textwidth]{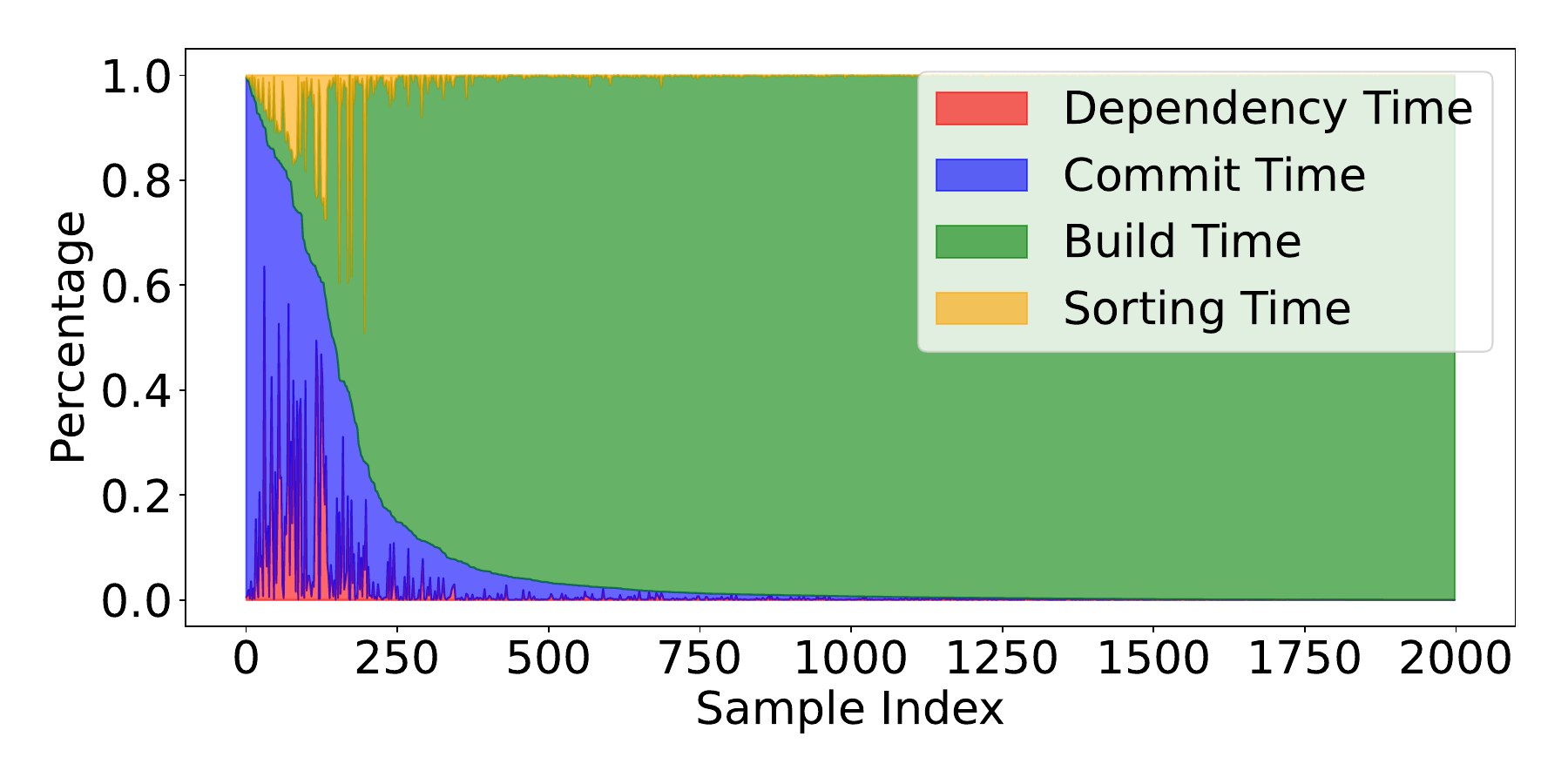}
        \caption{Time Cost of Each Step}
        \label{fig:rq1-4}
        \Description{Time Cost of Each Step}
    \end{minipage}%
    \hfill
    \begin{minipage}{0.48\textwidth}
        \centering
        \includegraphics[width=\textwidth]{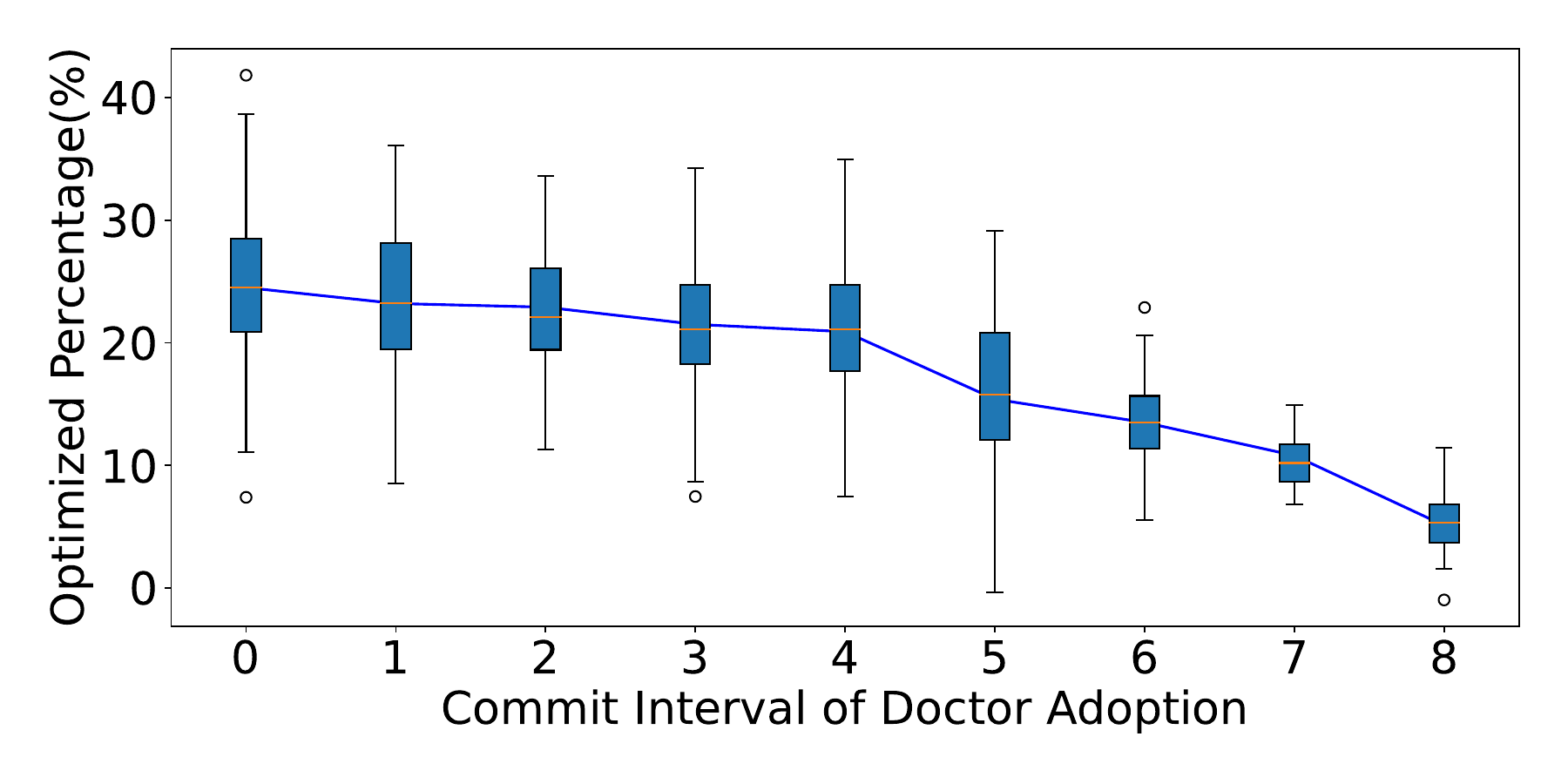}
        \caption{Effectiveness of Different Use Frequency}
        \label{fig:rq1-3}
        \Description{Effectiveness of Different Use Frequency}
    \end{minipage}
\end{figure*}

\noindent \textbf{\added{Efficiency Evaluation.}} \replaced{Regarding efficiency, \toolname took an average of 77.55 seconds to complete the optimization process for each Dockerfile.}{Regarding efficiency, \toolname required an average of 84.60 seconds to complete the optimization process for each Dockerfile.} As depicted in ~\Cref{fig:rq1-4}, most optimization time was allocated to the build time collection and commit collection step, highlighting the impact of "slow build" issues. The remaining steps were completed within an average of 3 seconds each. \added{Furthermore, in practical use, since each commit record retrieval only requires analyzing the incremental part, the time spent on this part will also be reduced.}

\noindent \textbf{\added{Using Frequency Evaluation.}} \added{Another important factor influencing the efficiency of \toolname in real-world scenarios is its usage frequency. In the effectiveness experiment, we optimized every version (commit) of the Dockerfile to calculate the optimization efficiency. However, in practical applications, such a high frequency of usage may not be necessary, and thus a trade-off between usage frequency and optimization results needs to be considered. To explore the balance between tool usage frequency and optimization effectiveness, we conducted further experiments. By varying the number of commits between optimizations, we observed the changes in the optimization results, as shown in \Cref{fig:rq1-3}. When the commit interval is set to 5, a significant decline in optimization efficiency is observed. Therefore, we believe that an interval of 4 versions (i.e., optimizing every 5 commits) strikes a balance between efficiency and optimization results. We calculated the average commit interval for all samples, with the average commit cycle of 5 commits being 3.7 months. Thus, in practical usage scenarios, using the tool every 3.7 months would be optimal.}

\begin{figure*}
    \centering
    \includegraphics[width=\textwidth]{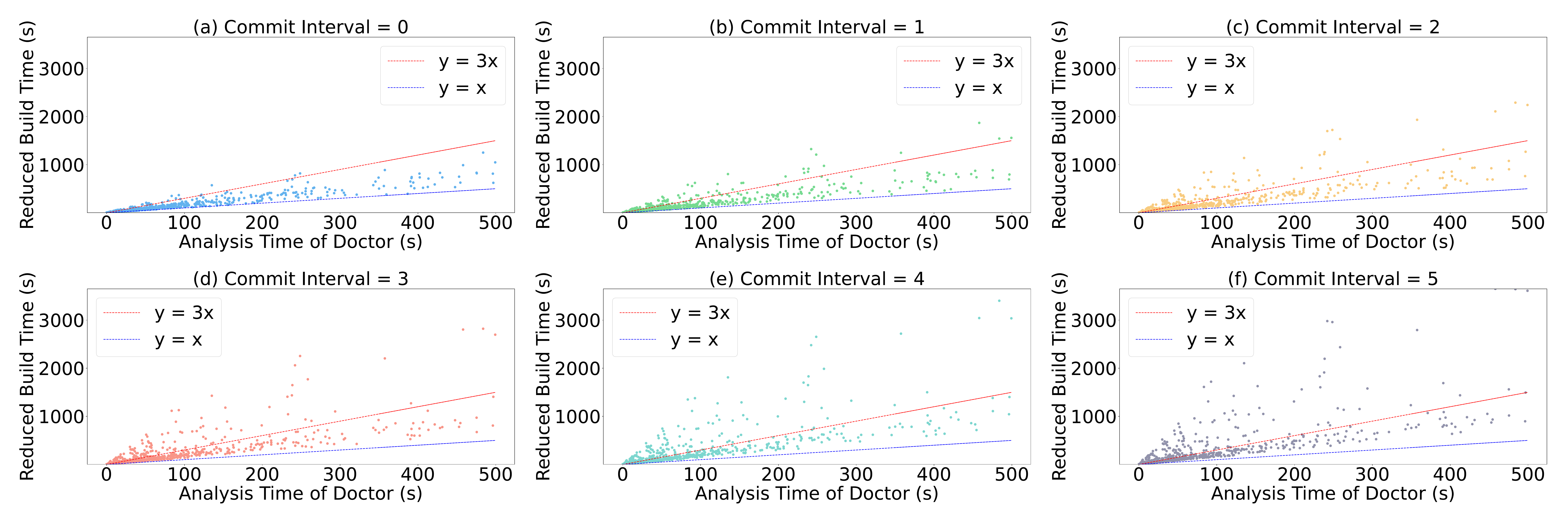}
    \caption{\added{Cost Time vs. Saving Time in Different Commit Interval}}
    \label{fig:rq1-9}
    \Description{Cost Time vs. Saving Time in Different Commit Interval}
\end{figure*}

\noindent \textbf{\added{Benefit-Cost Evaluation in Practice.}} \added{To further assess the cost-benefit ratio in practice, we conducted a detailed analysis of time savings, as shown in \Cref{fig:rq1-9}. Using the tool to optimize each modification, we achieved an average time saving of 110.75 seconds (calculated as the sum of time saved between consecutive optimizations), as depicted in the first sub-figure. This saving already exceeds the tool's time cost. When \toolname was used less frequently, as shown in the following sub-figure, the time savings generally outweighed the tool's cost. Specifically, when the tool was applied after every 5 Dockerfile changes, the average time savings reached 212.85 seconds, which is three times the cost of using the tool.}

\added{Furthermore, in practice, building containers based on the same Dockerfile can be much more frequent (i.e., regular CI/CD and duplicated instances in cloud services), by different roles (i.e., developers, users). We think the real benefit is far larger than only the time reduced in single builds and deployments. For instance, for the realm/realm-java \cite{realm} project, which is a mobile database with 11.5k stars and 1.8k forks, \toolname can averagely reduce the build time from 137s to 72s per build (47.4\% improvement), which could greatly improve the productivity of downstream users.}


\mybox{\textbf{Answer to RQ1.2:}
\toolname requires an average of 77.55 seconds to optimize a Dockerfile. Excluding the time required to build the Dockerfile itself, the optimization process is completed in approximately 3 seconds. In practical use, since only incremental data needs to be retrieved, this time will be further reduced. Regarding usage frequency, optimizing the Dockerfile every 3.7 months strikes the optimal balance between overhead and optimization efficiency. In real-world scenarios, the time saved typically offsets the tool's time consumption, and in peak cases, the benefit-cost ratio can reach up to three times.}


\subsection{RQ2: Consistency Analysis}

In this section, we evaluate whether Dockerfiles maintain functional equivalence following optimization by \toolname. As no existing method directly measures the functional equivalence of two Docker images, we employed several approximation metrics to assess this equivalence. 

\begin{itemize}[leftmargin=*]
    \item \textbf{File-system Structure}: This metric assesses whether the optimized Docker image maintains the same directory structure and file contents as the original. We use a recursive directory analysis with hash comparisons to detect any unintended changes in file integrity or organization.
    
    \item \textbf{Environment Variables}: This metric verifies that the set of critical environment variables remains consistent post-optimization. We retrieve all environment variables with \texttt{docker inspect} and exclude dynamically generated ones (e.g., \texttt{HOSTNAME}, \texttt{PWD}) to focus on those essential to the application’s functionality.
    
    \item \textbf{Package Manager Contents}: This metric checks that the list of installed packages is consistent before and after optimization. By confirming package consistency, we ensure that all necessary dependencies remain intact following instruction reordering.
    
    \item \textbf{WORKDIR}: This metric ensures that the default working directory, as specified in the Dockerfile, is preserved in the optimized image. The working directory serves as the entry point for container operations, so maintaining it is crucial for functional equivalence.

    \item \textbf{Unit Test (If Available)}: \added{If the repositories include unit tests or CI/CD-related tests, verify whether the optimization still passes the tests.}
\end{itemize}

\replaced{We conducted experiments on the latest version of each Dockerfile in the dataset with over 1,000 stars (i.e., repositories with high quality) and compared the differences before and after optimization.}{We conducted experiments on the latest version of each Dockerfile in the dataset and compared pre- and post-optimization differences.} Results showed that 86.2\% of images retained directory structures post-optimization, with no significant changes in file content. Among the 138 cases with content differences, the primary causes were discrepancies in the file-system structure and package manager contents (114 cases and 87 cases, respectively). Variations in environment variables and \texttt{WORKDIR} were minimal (13 cases and 0 cases, respectively).

\added{We conducted a manual inspection of the 138 cases with differences by checking the basic functionality of the optimized containers based on the README or description. (e.g. starting the mentioned service)} \replaced{The result}{Manual inspection of the 138 cases with differences} revealed that 84.8\% (117 cases) still maintained basic functional similarity despite minor discrepancies. These differences were primarily due to automatically updated dependencies in package managers (e.g., \texttt{apt} retrieving the latest versions) and dynamically generated files or configurations during container initialization. These variations did not impact core functionality but led to slight configuration changes. In the remaining 21 cases, differences arose due to unresolvable dependencies within complex \texttt{SHELL} instructions, resulting in certain dependencies failing to install post-reordering. These rare situations typically occur only in specialized use cases and thus do not warrant additional handling.

\added{As for the unit test verification, by filtering the repository's README and description, we identified 23 repositories with unit test cases from the dataset. Among the 23 repositories, 13 repositories are used to develop a basic CI/CD development pipeline, 7 repositories are used for testing and quality assurance, and 3 repositories are used as a template to quickly develop a container runtime. We manually re-ran the repository's unit tests in the optimized containers, and test cases from all 23 repositories passed successfully.}




\mybox{\textbf{Answer to RQ2:}
Our results show that \toolname maintains functional similarity in most cases. After optimization, 86.2\% of Docker images preserved consistency with the static environment. Manual analysis based on the README confirmed that 117 repositories retained functional similarity. Furthermore, all 23 repositories with unit tests passed post-optimization. Only 0.21\% of cases exhibited semantic differences, indicating that such occurrences are rare in practice.}

\subsection{RQ3: Ablation Study}

To understand the individual impact of each factor on optimization efficiency and build success rate, we designed an ablation study focused on three variables: Dependency, Modification Frequency (MF), and Build Cost (BC). We selected the latest version of each Dockerfile in the dataset and conducted separate experiments by systematically altering each variable and observing its effect on the final topology sorting outcome. For Dependency, we removed all dependencies between instructions and performed a new round of topology sorting. For MF, we set the modification frequency of all instructions to a uniform value, simulating an equal likelihood of changes across all instructions. Similarly, for BC, we assigned a uniform build cost to all instructions, eliminating differentiation based on individual instruction costs.

\begin{figure*}[t]
    \centering
    \includegraphics[width=0.8\textwidth]{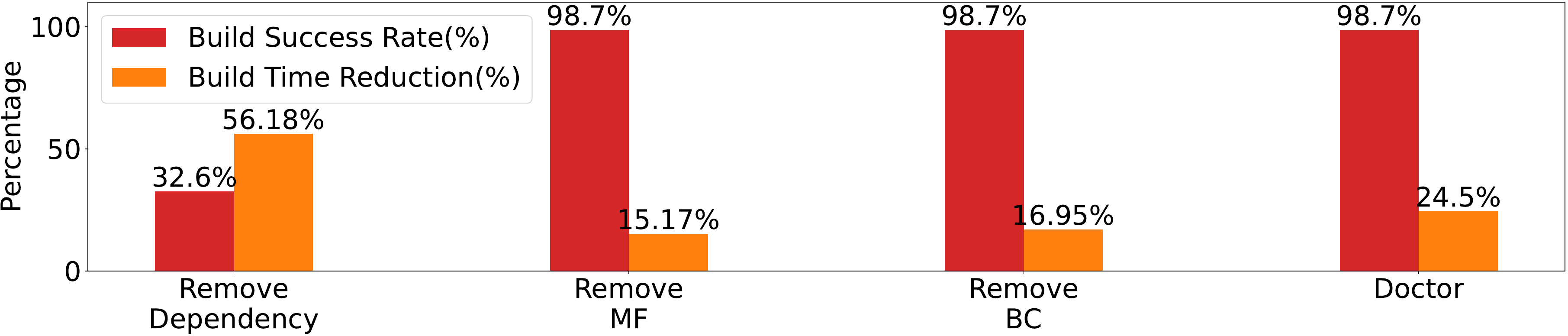}
    \caption{Success Rate of Builds and Optimization Percentage for Each Factor Removal}
    \label{fig:rq2}
    \Description{Success Rate of Builds and Optimization Percentage for Each Factor Removal}
\end{figure*}

\added{The ablation study compared the optimization results of each experiment with the original sequence. As shown in ~\Cref{fig:rq2}, removing dependencies significantly affected the build success rate, with 67.4\% of Dockerfiles failing to build correctly. In contrast, setting modification frequency and build cost to uniform values resulted in moderate improvements in optimization effectiveness. Specifically, adjusting modification frequency (MF) improved build efficiency by 7.55\%, while setting build cost (BC) to a uniform value improved it by 9.33\%.}

Through manual inspection of content differences between the original and ablation study builds, we observed that removing Dependency disrupted essential \texttt{FROM} relationships, which notably decreased the build success rate. Without these dependencies, critical base image instructions were reordered incorrectly, resulting in widespread build failures. In contrast, setting all modification frequencies to a uniform value primarily influenced frequently changing instructions, leading to a greater performance improvement than adjustments in build cost.

\added{To validate the reasonability of prediction future changes with historical records, we experimented on the modification prediction model. For each repository, we used the past 80\% of commits to predict the next modification probability distribution, selecting the highest probability as the prediction. 
Our experiment showed that the model achieved an average accuracy of 93.15\%, demonstrating its ability to capture correlations between past records and future changes. We further analyzed prediction accuracy for specific command types, finding that \textit{RUN} and \textit{COPY} instructions had the highest accuracies, 94.5\% and 92.3\%, respectively. Specifically, instructions related to installation and environment setup, such as APT and PIP commands (\textit{apt} 97.2\%, \textit{apt-get} 96.8\%, and \textit{pip} 94.8\%), were most accurately predicted. Similarly, commands for copying files to the current path (``COPY <file> .", 96.3\%), root path (``COPY <file> /", 96.1\%), or working directories (``COPY <file> /app", 95.2\%) also showed high prediction accuracy.}

\added{In contrast, we also identified some instructions with only very limited accuracy. For instance, \textit{ONBUILD} and \textit{STOPSIGNAL} instructions exhibited lower prediction accuracies of 32.7\% and 15.2\%, respectively. Specifically, ``ONBUILD RUN" and ``STOPSIGNAL SIGINT" reached the lowest accuracy, which were 5.2\% and 3.8\%, respectively.}

\added{These results showed that historical records can indeed reflect and predict their future modifications with high accuracy since the historical records have adequate data for prediction, while on less concerned instructions, the prediction is with poor accuracy. However, since the frequently modified instructions in historical data takes a significant portion, i.e., over 80\% of modifications are actually \textit{RUN} and \textit{COPY} instructions, the prediction, in most cases, are accurate, and our hypothesis that historical modifications provides a reliable indicator of the probability of future changes would work in most cases.}




\mybox{\textbf{Answer to RQ3:}
The ablation study results underscore the importance of Dependency in maintaining build success, as its removal disrupts the critical instruction order. Build Cost has a greater impact on optimization effectiveness than Modification Frequency, suggesting that prioritizing time-intensive instructions can improve build efficiency. Additionally, the prediction model achieved an accuracy of 93.15\%, and further analysis of specific command types assessed its effectiveness.}

\subsection{RQ4: Contribution Analysis}

\begin{figure*}[t]
    \centering
    \includegraphics[width=0.8\textwidth]{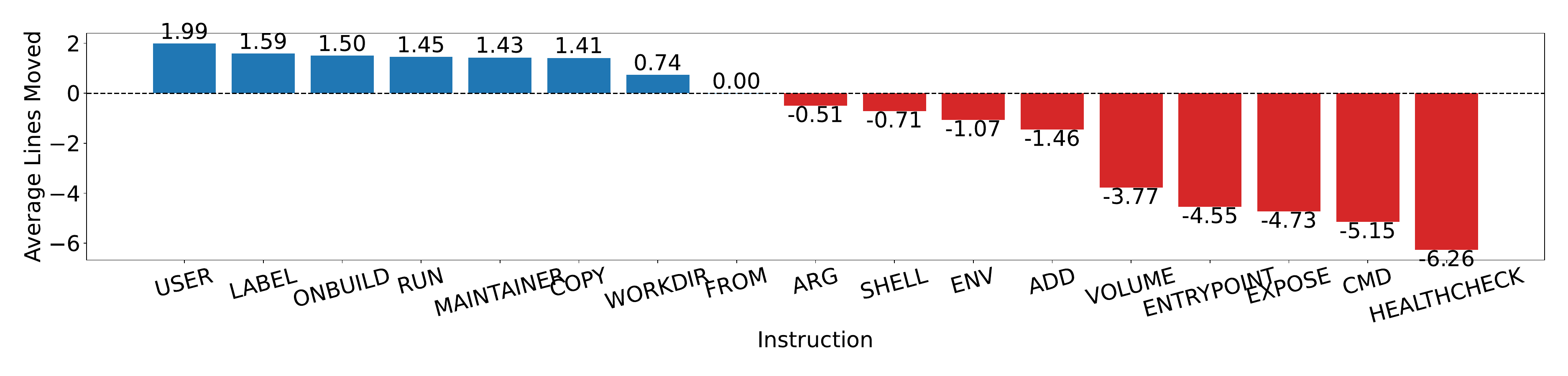}
    \caption{Average Number of Lines Moved of each Instruction}
    \label{fig:rq4-2}
    \Description{Average Number of Lines Moved of each Instruction}
\end{figure*}

To identify common patterns emerging from Dockerfile optimizations that contribute the most to \toolname, we first counted the average number of lines moved by each instruction before and after optimization, as shown in ~\Cref{fig:rq4-2}. This value indicates the change in line number after optimization compared to before optimization. If it is positive, it means that the instruction is post-positioned after optimization; otherwise, it is pre-positioned. 

For further mining, we manually inspected the optimization contents of repositories with over 1,000 stars. We aimed to uncover the primary objectives and methods behind each optimization and summarize the main considerations. The inspection followed these steps: First, three authors reviewed the Dockerfiles to identify changes and key considerations. Next, two authors reviewed and consolidated these observations into a preliminary conclusion. Finally, all authors discussed to ensure the conclusion was sound and representative. Following these steps, we identified several recurring patterns in the optimizations that could guide best practices in Dockerfile writing for achieving an optimal instruction sequence. The main optimization strategies are as follows:

    \noindent $\bullet$ \textbf{Positioning of FROM and WORKDIR (88.3\% of cases)}: The \texttt{FROM} instruction consistently appears at the start of the optimized sequences, underscoring its crucial role in facilitating efficient layer caching. Similarly, \texttt{WORKDIR} is generally positioned early in the sequence to establish the working directory, enabling subsequent instructions to leverage this setup.

    \noindent $\bullet$ \textbf{Forward Placement of ENV (82.5\% of cases)}: Placing \texttt{ENV} commands in the initial stages of the Dockerfile ensures that environment variables are defined before other instructions, maximizing layer reuse by minimizing reconfiguration later in the process.

    \noindent $\bullet$ \textbf{Postponing Installation Commands (76.2\% of cases)}: Package installation commands, such as \texttt{RUN apt install} or \texttt{RUN pip install}, are moved towards the end of the optimized Dockerfile. This approach helps isolate frequently changing instructions, ensuring that foundational layers are cached effectively while only volatile commands are rebuilt as needed.

    \noindent $\bullet$ \textbf{CMD Positioning (67.8\% of cases)}: The \texttt{CMD} instruction is often moved earlier in optimized Dockerfiles to stabilize the container’s final state early, reducing the frequency of rebuilds.

Based on the optimization pattern obtained above, we re-examined how developers modified the Dockerfile instructions in practice. Most modifications are caused by \texttt{COPY} (71.39\%) and \texttt{RUN} (23.32\%) instructions. To better utilize the Docker cache, they should be postponed. For the rest of the instructions, the modification portion is less than one percent. These not usually changed (i.e., \texttt{CMD, ENV}) instructions should be put forward to minimize the influence of the following instructions. This practice provides good support for optimization patterns.

During the inspection, we observed that developers often grouped related instructions for readability and ease of future modification. For instance, setting up \texttt{JDK} environment typically involves multiple instructions, including updating \texttt{apt}, defining the \texttt{JAVA\_HOME} path, and installing the \texttt{JDK} package via \texttt{apt}. These instructions are grouped for readability, facilitating quick modification. However, after optimization, these semantically related groups may be separated to maximize efficiency. To further investigate the relationship between readability loss and efficiency gain, we randomly selected 100 samples from the dataset for analysis. Five authors manually reviewed the optimized instruction sequences of each sample and then reordered instructions to restore semantic grouping. Since semantic interpretation can vary, we averaged the results from the manual review. The \toolname optimization increased the average build efficiency of these samples by 26.3\%. After restoring semantic groupings, the efficiency improvement dropped to 18.4\%. This indicates a trade-off between readability and build efficiency, which developers need to weigh carefully.

\mybox{\textbf{Answer to RQ4:} 
Our analysis revealed four patterns of optimizations, offering a framework that effectively leverages caching to reduce rebuild times. However, reordering may compromise readability and increase maintenance costs. The trade-off between efficiency and readability is approximately 5.9\%, which developers should consider in their design.}

\section{Discussion}
\label{sec:discussion}

\noindent $\bullet$ \textbf{Trade-off between Efficiency and Readability}. \replaced{One of the most significant challenges in optimization was balancing the trade-off between efficiency and readability. Developers often grouped instructions related to the same functionality together, despite differences in modification frequencies. This structuring enhances the semantic clarity of the Dockerfile, making it easier for developers to maintain and modify. However, during optimization, \toolname reordered these instructions, which disrupted these semantically cohesive blocks. While this resulted in notable improvements in build times, it also decreased the readability of the Dockerfile. In practice, developers may prioritize a more readable and modifiable Dockerfile over optimal build efficiency. This tension between theoretical optimization and practical usage is a key consideration for future work. To increase adoption, future optimization tools should offer customization options that allow developers to balance optimization with the specific needs of their projects and team dynamics.}{In our optimization process, one of the most significant challenges was balancing the trade-off between efficiency and readability. We observed several cases where developers grouped instructions related to the same functionality together, even though these instructions had different modification frequencies. This structuring of instructions enhances the semantic clarity of the Dockerfile, making it easier for developers to maintain and modify in the future. However, during optimization, our tool reordered these instructions to maximize build efficiency, disrupting these semantically cohesive blocks. While this led to notable improvements in build times, it simultaneously decreased the readability of the Dockerfile. In practice, developers may prefer Dockerfile structures that are easier to read and modify, even if that means sacrificing some level of build efficiency. This tension between theoretical optimization and practical usage is an important consideration for future work. To increase adoption, future optimization tools might need to offer customization options that allow developers to fine-tune the level of optimization, balancing it with the specific needs of their project and team dynamics.}

\noindent $\bullet$ \textbf{Dockerfile Modification and Project Evolution}. \added{As projects evolve, Dockerfiles inevitably require updates to accommodate new dependencies, changes in the build process, or modifications to the application environment. These updates, while necessary, can introduce complexity and inconsistency, especially in larger projects where frequent modifications are common. Currently, there is no standardized or universal approach to managing these changes, leading to Dockerfiles that can become unwieldy and difficult to maintain over time. One potential solution is to decompose the Dockerfile into smaller, more manageable sub-modules. By abstracting common configuration steps into reusable templates, we can isolate frequently modified sections, ensuring that only the affected sub-modules are updated. This modular approach not only reduces the burden of ongoing maintenance but also promotes the use of shared templates for common configurations.}


\noindent $\bullet$ \textbf{Long-term Maintenance vs. Short-term Performance Gains}. 
Dockerfile optimizations often focus on immediate performance gains by reducing build times. However, these short-term improvements may introduce longer-term maintenance challenges. As Dockerfiles become more optimized, they can also become more complex and harder to understand and modify. Over time, the increasing complexity of an optimized Dockerfile could lead to a situation where future developers face difficulty maintaining it. This introduces a trade-off between achieving quick performance gains and ensuring long-term maintainability. In the long run, the cost of maintaining an overly optimized Dockerfile might outweigh the short-term performance benefits. This raises the question of how to strike the right balance between optimization and maintainability over the lifecycle.

\noindent $\bullet$ \textbf{\added{Future Work Directions.}} \added{One of the future directions for enhancing Dockerfile build efficiency is incorporating smell remediation techniques into the re-orchestration optimization strategy that improves both quality and build performance. By integrating smell detection into the dependency extraction process, container images can be streamlined without disrupting essential dependencies. Another promising avenue is on-the-fly re-orchestration, where the optimization takes place exclusively during the build process. This enables real-time mapping of the Dockerfile and existing build cache, eliminating the reliance on historical data and allowing direct optimization based on the current cache. Furthermore, systematic techniques for defining and characterizing container behavior can facilitate more comprehensive dynamic comparisons between the original and optimized Docker environments, thereby advancing the best evaluation in Docker image optimization.}
\section{Threats to Validity}
\label{sec:threats}


\noindent $\bullet$ \textbf{Dependency Extraction}. The predefined rules might oversimplify certain dependencies that span across multiple layers of the Dockerfile, such as complex instructions like multi-stage builds or advanced shell commands, and dependencies on external resources may not be completely captured. To deal with this, we have systematically inspected the official documentation of Dockerfile, and proposed the EBNF to facilitate the extraction, we have also introduced libdash~\cite{libdash} to parse the embedded shell commands to detect and link all elements as much as possible.

\noindent $\bullet$ \textbf{\added{Rebuild Time Variability and Platform Constraints}}. \added{Rebuild time measurements can be affected by system performance variability, including network latency, hardware limitations, or operating system differences. To mitigate this, we repeated the build process three times with cleanups and used the average as the reliable build time. Additionally, the optimization method depends on cache management and multiple builds, which can be influenced by platform factors, such as computing resource availability. To ensure robustness, we conducted a large-scale experiment on 2,000 repositories from diverse ranges.}


\noindent $\bullet$ \textbf{\replaced{The Bias between Future Changes and the History.}{Matching Strategy for Modification Frequency}} \added{Since the motivation for the coming modification can be due to different reasons, the past records cannot directly reflect the probability distribution. Although, the prediction accuracy is 93.15\%, it can still be improved.} \added{Besides,} the matching strategy used to estimate modification frequency (i.e., strict matching, similarity-based matching) could introduce potential inaccuracies. To mitigate this, we adopted a hybrid strategy to design matching rules for different types of instructions to reduce possible inaccuracy. 


\noindent $\bullet$ \textbf{\added{Functional Consistency Evaluation.}} \added{Containers host software with diverse functions, making it challenging to determine if their behavior meets expectations. While environmental parameters can provide some insight, effective methods to characterize dynamic container behavior remain lacking. To address this, we use unit tests within the container as a verification standard, combined with static environment parameters, reducing verification ambiguity.}

\section{Related Work}
\label{sec:relatedwork}

\noindent $\bullet$ \textbf{Build Script Optimization and Refactoring.} \added{Build script optimization has focused on enhancing the manageability, performance, and reliability of traditional systems. For instance, Gligoric et al. ~\cite{gligoric2014automated} introduced Metamorphosis for migrating legacy scripts to modern systems, improving parallelization and maintainability, while Tamrawi et al. ~\cite{tamrawi2012symake} refined Makefiles by addressing code smells and cyclic dependencies. Other studies, such as Macho et al.'s BUILDDIFF~\cite{macho2021nature}, track build script evolution, aiding automated repairs by analyzing version changes. Research has also explored the impact of refactoring on non-functional aspects, as Traini et al. ~\cite{traini2021software} found potential performance regressions. Tools like HireBuild ~\cite{hassan2018hirebuild} and Gradle-AutoFix~\cite{kang2022gradle} automate Maven and Gradle script repairs. However, while these advancements improve traditional build scripts, they fail to address the unique challenges of Dockerfile optimization, which involves image caching, dependency sequencing, and build frequency, necessitating specialized research in this area.}

\noindent $\bullet$ \textbf{Docker Performance.} Recent research on Docker performance has mainly focused on layer size, efficiency, and code quality, with limited attention to sequence optimization and modification frequency. Wu et al.~\cite{wu2022understanding} found that inefficient builds impede productivity and developed a predictive model using 27 features to reduce build times. Studies on Dockerfile quality emphasize reproducibility challenges. Cito et al.~\cite{cito2017empirical} highlighted issues like missing version pinning and suggested structured abstractions and lightweight images for improved reliability. Henkel et al.~\cite{henkel2021shipwright} introduced SHIPWRIGHT, an automated tool that outperforms traditional static analysis in detection and repair. Dockerfile refactoring has also gained attention, with Ksontini et al. ~\cite{ksontini2021refactorings} identifying practices to optimize image size, and DRMiner ~\cite{ksontini2024drminer} detecting refactoring candidates using Enhanced Abstract Syntax Trees. Huang et al.'s FastBuild ~\cite{huang2019fastbuild} improved speed by caching remote file requests, increasing build performance by up to 10x. Studies by Zhao et al. ~\cite{zhao2020large} and Durieux ~\cite{durieux2024empirical} focused on storage efficiency and reducing Docker smells, with an emphasis on layer size rather than sequence optimization. Despite these advancements, research has not explored the impact of reordering Dockerfile instructions or considering modification frequency to optimize cache efficiency, indicating a need for further investigation into these factors.

\section{Conclusion}
\label{sec:conclusion}

This paper introduced \toolname, an approach to optimizing Dockerfile build efficiency by restructuring instructions based on dependency analysis and modification patterns. Tested on \replaced{2,000}{1,000} repositories, \toolname achieved an average \replaced{26.5\%}{24.5\%} reduction in rebuild time. Additionally, we identified four optimization patterns to guide Dockerfile development. By open-sourcing our dataset on Dockerfile dependencies, we provide a foundational resource for further research in containerization, which offers an effective solution for balancing build efficiency and development demands.

\section*{Acknowledgment}
This research is supported by National Natural Science Foundation of China under Grant No. 62002324, U22B2028, and U1936215. Science and Technology Projects of Zhejiang Province - High-Level Talents Special Support Program under Grant No. 2020R52011, the Ministry of Education, Singapore, under its Academic Research Fund Tier 1 (RG96/23). It is also supported by the National Research Foundation, Singapore, and DSO National Laboratories under the AI Singapore Programme (AISG Award No: AISG2-GC-2023-008); by the National Research Foundation Singapore and the Cyber Security Agency under the National Cybersecurity R\&D Programme (NCRP25-P04-TAICeN); and by the Prime Minister’s Office, Singapore under the Campus for Research Excellence and Technological Enterprise (CREATE) programme. Any opinions, findings and conclusions, or recommendations expressed in these materials are those of the author(s) and do not reflect the views of National Research Foundation, Singapore, Cyber Security Agency of Singapore, Singapore.

\section*{Data Availability}
The dataset and the code of \toolname are published in \cite{home-page}.

\clearpage

\bibliographystyle{ACM-Reference-Format}
\bibliography{acmart}

\end{document}